\documentclass{article}

\title{Network Inference from a Link-Traced Sample using Approximate Bayesian Computation}
\date{December 29, 2016}
\author{Jack Davis\\ Simon Fraser University \and Steven K. Thompson\\ Simon Fraser University}

\usepackage{subfigure_fix}
\usepackage{setspace}

\usepackage{amsmath,amssymb,amsthm}
\usepackage{graphicx}
\usepackage[]{algorithm2e}
\usepackage{multirow}

\begin{document}
\begin{titlepage}
\maketitle

\begin{abstract}

In this manuscript, we present a new inference method based on approximate Bayesian computation for estimating parameters governing an entire network based on link-traced samples of that network. To do this, we first take summary statistics from an observed link-traced network sample, such as a recruitment network of subjects in a hard-to-reach population. Then we assume prior distributions, such as multivariate uniform, for the distribution of some parameters governing the structure of the network and behaviour of its nodes. Then, we draw many independent and identically distributed values for these parameters. For each set of values, we simulate a population network, take a link-traced sample from that network, and find the summary statistics for that sample. The statistics from the sample, and the parameters that eventually led to that sample, are collectively treated as a single point. We take a Kernel Density estimate of the points from many simulations, and observe the density across the hyperplane coinciding with the statistic values of the originally observed sample. This density function is treat as a posterior estimate of the paramaters of the network that provided the observed sample.

We also apply this method to a network of precedence citations between legal documents, centered around cases overseen by the Supreme Court of Canada, is observed. The features of certain cases that lead to their frequent citation are inferred, and their effects estimated by ABC. Future work and extensions are also briefly discussed.

\end{abstract}
\end{titlepage}

\section{Introduction}

Link-tracing samples (sometimes called respondent-driven samples or snowball samples in social science contexts), or adaptive web sampling (Thompson 2006)\nocite{thompson2006adaptive} are used to sample hard-to-reach networked populations, such as endangered species, injection drug users, or other at-risk people. They are commonly used in the social sciences, and are a large source of non internet-based social network data.

In a link-tracing sample, part of the target population is selected, ideally by simple random sampling, but often by convenience or directed sampling. This set of initial contacts is sometimes called the `seeds' of the sample. The variables of interest are measured from these seeds, and some mechanism is used to find additional subjects that are connected to at least one of the seeds in some predefined way. Common mechanisms are to ask seeds for contact information of their connections, and/or to give recruitment coupons to seeds that can be given to connected population members to entice these population members to join the sample and be measured as well.

The first wave of connected population members that are sampled are then asked for contacts and/or given coupons as the seeds were in order to bring a second wave of members into the sample. Sampling continues in this fashion until no new members of the population are recruited into the sample, or a predetermined sample size has been reached. Additional seeds may be selected if recruitment is exhausted before the sample size is reached.

As a proportion of published research using link-traced samples, there is relatively little focus on the network structure of the respondents. There are many barriers to inference of the network of a population from a link-traced sample. Stopping after a fixed sample size, non-response, non-recruitment, and selection of the initial seeds to recruit from all affect the sample network that is observed.  For example, in some cases, connection information that leads to subjects already in the sample is discarded because it cannot bring in new respondents.  

Use of link-traced samples in research is well established (Goodman 1961), including in the context of sampling from hard-to-find populations (Kaplan, Korf, and Sterk 1987), and with the respondents themselves determining recruitment (Heckathorn 1997). \nocite{goodman1961snowball} \nocite{kaplan1987temporal} \nocite{heckathorn1997respondent}  Advances in finding the network structure of the respondents is relatively recent. Crawford (2014) \nocite{crawford2014graphical} describes how partial information about a network structure can be inferred from a respondent-driven sample. Hancock, Gile, and Mar (2015) applies a respondent-driven sample to a population at risk of HIV infection, and infers features of the network including the number of nodes (i.e. the order of the network). \nocite{handcock2015estimating} 

There are many barriers to inference of the network of a population from a link-traced sample. Stopping after a fixed sample size, non-response, non-recruitment, and selection of the inital seeds to recruit from all affect the sample network that is observed.  For example, in some cases, connection information that leads to subjects already in the sample is discarded because it cannot bring in new respondents.

Inference of the specific structure of a population network may be infeasible from a snowball sample. However, it is possible to infer general features of such a network, such as the distribution of the number of connections between respondents, and parameters that dictate which connections form.

Section 2 describes the approximate Bayesian computation (ABC) method in a general context. This section also contains a literature review of some recent advances in ABC that are closely related to the proposed method.

Section 3 includes a short glossary of network terms being used. It follows with some example network structure-based statistics established in previous work, and some sampling order-based statistics that are established specifically for this method.

 Section 4 describes the method being proposed. This includes an expansion upon the description of ABC from Section 2 to apply it to a snowball sample. Also included are the steps taken to randomly generate the population network necessary for this method, the protocol used for taking a snowball sample, and the kernel density estimator used to estimate the probability density of the parameters given the sample's statistics.

Section 5 gives a demonstration of the proposed method on snowball samples taken from some populations with known parameters. In the demonstration, the ideal sample statistics are unknown, so two rounds of the network inference methods are performed. The initial round uses a small simulation run to determine which summary statistics of the sample would be useful, and to update the parameter priors. The final round uses a large simulation run and an updated prior to give a high-resolution estimate of the posterior of the parameters. Posterior means are taken from the final round estimate. Results for both rounds are given as topographic maps of the conditional probability density. 

Section 6 provides the motivation for and outline of the application of the proposed method to the CanLII database of Supreme Court of Canada decisions.

Section 7 describes the application in greater detail and provides results.

Section 8, the conclusion, describes how this method can be adapted to applications with additional parameterization. It also provides some priorities for future development of the method and its software to address wider application and computational challenges.

\section{Recent Advances in approximate Bayesian computation}

Approximate Bayesian computation (ABC) is a Monte Carlo-based method of estimating the distribution of parameters from which inference would otherwise be intractable. It is used extensively in describing complex systems in evolutionary biology (Csill{\'e}ry et al. 2010)\nocite{csillery2010approximate}, non-linear regression models in statistical genetics (Blum and Fran{\c{c}}ois 2010)\nocite{blum2010non}, and agent-based models in oncology (Sottoriva, and Tavar{\'e} 2010)\nocite{sottoriva2010integrating}.

To perform a classic ABC on a sample, as described in Diggle (1984)\nocite{diggle1984monte}, first calculate statistics $s'$ from the data, define a distance function $\rho(s,s')$ such as Euclidean distance, and define a tolerance parameter $\epsilon$. In a classic ABC, parameter values are randomly generated from a prior, and datasets are generated according to a given model and these parameters.

For each of the randomly generated datasets, the same statistics $s$ are taken as those that were taken from the original data. If the distance between $s$ and $s'$, $\rho(s,s')$ is greater than $\epsilon$, the parameter set used to generate this dataset is rejected. Otherwise, the parameter set is accepted. 


The sets of parameter values that are accepted are values that were used to produce simulated samples similar to the observed sample. We take this distribution (or some smoothing of it) to be the posterior distribution of parameter values.

Two common modifications to approximate Bayesian computation are partial rejection and adaptive parameter selections. When partial rejection is used, datasets are either accepted with probability or with weights, rather than accepted or rejected outright. The acceptance probability or weight is based on their statistics' distance to the target statistics. The weights of the simulations used in Beaumont et al. (2002) are determined by an Epanechnikov kernel, with density $\frac{3}{4} \left( 1 - d^2 \right)$ for distances $0 \leq d \leq 1$, and density $0$ otherwise. When adaptive parameter selection is used, the parameters that are used for each simulation are informed in some way by the results of the previous simulations, usually in order to produce simulations that have summary statistics close to the observed statistics. Parameters in the ABC implementation in Del Moral et al. (2012)\nocite{del2012adaptive} are adaptively improved with sequential Monte Carlo, and the criterion for rejection is made stricter as more simulations are used.

The proposed method employs partial rejection, but unlike the application of kernel density in Beaumont et al. (2002)\nocite{beaumont2002approximate}, the probability mass across a hyperplane is computed, rather than at a single point.  Some possibilities for adaptive parameter selection are discussed in Section 6.

Inference on network structure is a new and sparsely explored avenue of research. Mukherjee and Speed (2008)\nocite{mukherjee2008network} describes a Monte Carlo Markov-chain method of inferring the existence of specific edges in directed network graphs, but only for graphs of order (i.e node count) $\approx10$ and size (i.e edge count) $\approx20$. Related methods that address larger networks, like those described and proposed in K{\"u}ffner et al. (2012)\nocite{kuffner2012inferring} produce summary statistics with large amounts of uncertainty and noise, as demonstrated in Petri et al. (2015)\nocite{petri2015addressing}.

Applications of approximate Bayesian computation to network data are also rare. Toni et al. (2009)\nocite{toni2009approximate} briefly mentions the possibility of extending the ABC approach to dynamical systems therein to networks of chemical signals. The approach to find the size of a hidden population in F{\'e}lix-Medina and Thompson (2004)\nocite{felix2004combining} uses a link-tracing design and a simulation of many networks, but does not employ ABC. Phillips et. al. (2013) \nocite{phillips2013increased} use inference methods similar to ABC to model an HIV epidemic, but do not use the words `approximate' or `Bayesian' to describe their method.

The proposed method is very similar to Fay et al. (2014)\nocite{fay2014graph} in that it uses ABC, an observed network, and an assumed network model to make inferences about the parameters behind the generation of that network. However, in Fay et al. (2014), the observed network is assumed to be generated directly from the given model and parameters. For the method proposed herein, only a sample of the generated network is observed, rather than the entire network. Inferences are to be made on both the parameters that generated the population network as well as the parameters such as response-to-recruitment chance that dictate the sample network.

\section{Network terminology and metrics}

\subsection{Quick glossary of network terms}

Terminology in network analysis has not yet been standardized, so we chose to adhere to the terms used in Kolaczyk (2009)\nocite{kolaczyk2009statistical} and Kolaczyk and Cs{\'a}rdi (2014)\nocite{kolaczyk2014statistical}. That is, a network graph $<V,E>$ is an abstract structure composed of a set of $n$ nodes $v_i$ for $i=1, \ldots ,n$ $V$, and a set of edges $E$. Each edge in $E$ an ordered pair $(v_i, v_j)$, i,j members of $1, \ldots ,n$, representing a one-way connection from $v_i$ to $v_j$. If $i=j$, then the edge is a connection from $v_i$ to itself and the edge is called a self-loop, or `loop' for short.

The size of a network graph refers to the number of edges, contained in $E$. The order of a network graph refers to the number of nodes  in $V$. The size and order will be referred to as $N_E$ and $N_V$ throughout this manuscript.

If for a given edge $(v_i, v_j)$, there is an edge $(v_j, v_i)$, the connection between nodes $v_i$ and $v_j$ is considered to be bi-directional. If for all edges $(v_i, v_j)$, there exists an edge $(v_j, v_i)$, the network graph is called undirected; otherwise it is called a directed graph. If there exists more than one edge $(v_i, v_j)$ for any pair i and j, the network graph is called a multigraph.

A subgraph $<V^*, E^*>$ of $<V,E>$ is a network graph in which $V^*$ is a subset of $V$, or $E^*$ is a subset of $E$, or both. For all edges $(v_i, v_j)$ in $E^*$, at least one of $v_i$ or $v_j$ must belong to $V^*$. If both $v_i$ and $v_j$ are in $V^*$ for all edges in $E^*$, and all such edges in $E$ are in $E^*$, then the subgraph is called an induced subgraph.

A path from nodes $v_k$ to $v_l$ is a set of edges$\{ (v_k, v_i), ( v_i, v_m), \ldots , (v_n, v_j), (v_j, v_l) \}$, $i,j$ in $1\ldots n$. This number of edges in the set is called the length, and it is possible for a path to be of length 1. If $k=l$, then the path is called a cycle.

A connected component, or `component' for short, is a subgraph of a network graph in which, for each node in the component, there is a path either to or from each node in the component. If there is path both to and from each node, the component is strongly connected. Every component in an undirected network graph is strongly connected, and in the context of undirected network graphs, components are simply referred to as connected. A connected component that does not contain any non-trivial cycles (cycles involving 3 or more nodes) is called a tree. Also, if every node from the network graph that could be in a component already is, that component is considered maximally connected.

The method pertains to populations described by undirected graphs that are not multigraphs and do not have self-loops. The population may contain cycles, but the samples do not. Furthermore, since most samples considered do not include every node in the population, components of the network observed from the sample may not be maximally connected.

When the desired number of nodes have been sampled, any remaining recruitment links are ignored. This makes the snowball sample taken by this protocol a subgraph of an induced subgraph of the population network graph. That is, the nodes of the sample are a subset of the population nodes, and only links strictly between nodes with those subset are included in the subgraph. There are no links `leading out' of the sample.

\subsection{Limitations of structure-based network statistics of link-traced samples}

Completely observed network graphs can be summarized using statistics that are based solely on the structure of the graph, such as the distribution of edges per node, the distribution of component sizes, and more complex measures based on cliques and centrality. Such is the case with the method proposed in (Fay et al. 2014). However, when sampling is introduced such as the link-tracing protocols that are used in our method, many of these statistics lose their inferential utility.

For the link-tracing sampling protocol of interest, only edges that lead to recruitment are retained in the dataset. This means that no connections from respondents to other respondents already in the sample are retained. Thus, no non-trivial cycles that are in the sample are observed. The observed data shows only a subtree of each sampled network component.

In cases where only a subtree or subtrees of the network graph are available, many common measures of network structure fail to be useful.  Consider that the size (number of links) of a tree-shaped component is one less than the order (number of nodes). The network graphs observed from a subtree sample of $n$ nodes can only have average degree on $[0, 2(n-1)/n ]$. A graph with average degree 0 would be one with no links between any of the nodes. A graph with the maximal average degree of $2(n-1)/n$ would have all $n$ nodes in a single giant component. To see this, consider the construction of a recruitment network with one component from an arbitary starting point. To add a node to the network, we need to add an edge to an existing node to maintain the one-component property. Edges are only ever added when nodes are added, therefore the number of edges added is always equal to the number of nodes added. Now consider that a single node is a (trivial) recruitment network of one component. By induction, a recruitment network has one more node than edges. Since any edge contributes two degrees to the nodes, a one-component recruitment network of $n$ nodes has $2(n-1)$ total degrees. Obviously, recruitment networks with more components and $n$ nodes have fewer than $n-1$ edges.

More detailed methods of describing of network structure have similar problems. Consider a motif census. A motif is a small induced subgraph of larger network graph. The four possible non-isomorphic motif arrangements of order 3 and some of the 11 such order 4 arrangements are shown in Figure 1. A census of the order k motifs of an order $n_{samp}$ graph counts the number of motifs that appear in all ${ n_{samp} \choose 3}$ induced subgraphs of the order $n_{samp}$ graph.

\begin{figure}[ht]
\begin{center}
\includegraphics[angle=0,scale=0.70]{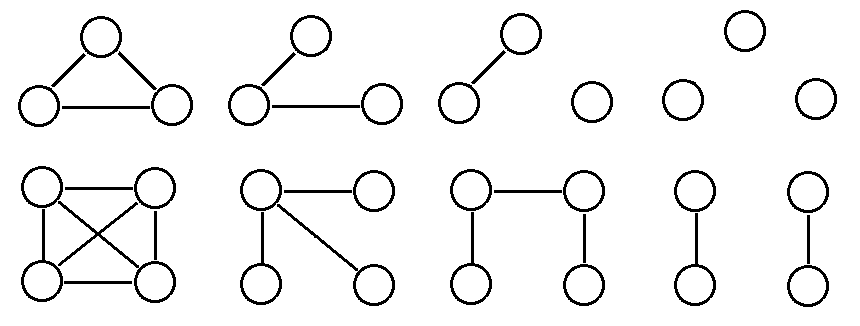}
\end{center}
\vspace{-8.0mm}
\caption{ Network motifs of order 3, and selected motifs of order 4}
\end{figure}

The counts from a motif census can be compared to the expected counts from networks of known random processes. The actual and expected motif counts can be compared with goodness-of-fit tests to infer if a given network could have been generated by a given process. 

When the graph is a single tree, as it is in cases where average degree fails to be useful, only a thin range of motif distributions are possible. Only considering motifs in a vacuum, only 3 of 4 order 3 motifs, and 6 of 11 order 4 motifs are possible without creating a cycle.

In addition to the above issues, other problems relating to non-response can arise. For example, each sample unit reports its own degree within the population, which may be greater than the degree observed in the sample network for one of three reasons: A node never responds to a given recruitment attempt, sampling is completed before a potentially recruited node responds, or a node is already in the sample, and does not respond after the first recruitment. In the face of these difficulties from relying solely on network structure to inform our summary statistics of choice, we also incorporate sampling order.

\subsection{Sampling-order based metrics}

We assume that for any sample, the chronological order in which nodes are added in a sample is available. Sampling order could be obtained directly by incrementing identity number, or inferred from time stamps.

The sampling order of any node affects the reasons that its edges may fail to be included in a sample. Edges from a node that are added near the end of a sample are less likely, all else being equal, to be followed because the desired number of nodes will have been sampled before these edges are followed. Similarly, as sampling continues, an increasing number of nodes are included, and, all else being equal again, a given link from a newly added node is more likely to lead to a node already in the sample.

The trend of increasing loss-to-redundancy is of special interest because it depends on the structure of the population network. Specifically, this trend occurs in samples taken from populations where there are more links than nodes, and is stronger in populations with higher link density, up to a given point. This is intuitive when one considers that in a sparsely connected graph, connected components will be small and will be explored completely in a relatively small part of a sample; leaps to new, unconnected units by simple random sampling are common, and each leap brings fresh connections. However, in network graphs with average degree between 2 and 6, a sample will typically stay in one component during its entire run, and such leaps to new components are not made. In densely-connected graphs where average degree is close to the order (i.e. where there are nearly $N_E = N_V(N_V-1)/2$ edges, in a graph with $N_V$ nodes) of the graph, links are so abundant that recruitment sampling begins to resemble simple random sampling in that the most recently sampled node gives little information about the next node to be sampled.

We compute $Pr_{LinkUsed}$, the proportion of a unit's reported links that are used for recruitment in a sample are defined by.

\vspace{5mm}

$Pr_{LinkUsed} =$ Recruitment links included in the sample $/$ Recruitment links reported.

\vspace{5mm}

If the node was recruited into the sample by another node, then

\vspace{5mm}

$Pr_{LinkUsed} = (|$Edges leading to sample$|- 1) / (|$Edges$| - 1)$.

\vspace{5mm}

If the node was selected from the population by simple random sampling, then

\vspace{5mm}

$Pr_{LinkUsed} = (|$Edges leading to sample $| - 1) / (|$Edges$|)$.

\vspace{5mm}


We define the measure $\Delta used / \Delta Sample$ to be the rate of change of the proportion of potential links used for recruitment over the time the sample is taken. We compute $\Delta used / \Delta Sample$ as the slope-coefficient of the linear regression model of $Pr_{LinkUsed}$, weighted by degree and as a function of $t$, where $t=0$ for the first unit sampled, and $t=1$ for the last unit. In other terms, $\Delta used / \Delta Sample$ is the estimated difference in proportion of recruitment attempts that are successful from the beginning to the end of a sample. This interpretation of the time variable $t$ is just one of several viable ones. For example, $t$ could refer to the number of nodes explored in each nodes' component, which would account for `leaps' in the link-tracing design to new components, but potentially lend undue weight to results from large components. Similarly, a binomial generalized linear model with a logit link could be used to find the ratio of the log-odds of recruitment at the beginning or the end of sampling. However, results based on the log-odds ratio becomes unstable for samples that use most or all of the nodes in the population because the chance of recruitment drops to near zero. A more philosophical limitation of $\Delta used / \Delta Sample$ is its blindness to the reason a given edge was used for recruitment, so whatever it reveals could be a result of non-response or edges being exhausted, and it isn't a panacea to the difficulties described in Section 2.

The panels of Figure 2 show an experimental example of how sample link density and how $\Delta used / \Delta Sample$, respectively, change as the average responding degree in a population increases. The metric $\Delta used / \Delta Sample$ changes rapidly as average population degree changes from 1 to 3 edges per node. At approximately 2.3 edges per node, the loess-smoothed average of $\Delta used / \Delta Sample$ reaches its minimum of $-0.4$, indicating that the recruitment chance at the end of a sample of one of these populations is 40 percentage points lower. This behaviour indicates that $\Delta used / \Delta Sample$ is useful in making inferences about networked populations in situations where metrics that ignore sampling order are not useful.

\begin{figure}[ht]
\begin{center}
\includegraphics[angle=0,scale=0.40]{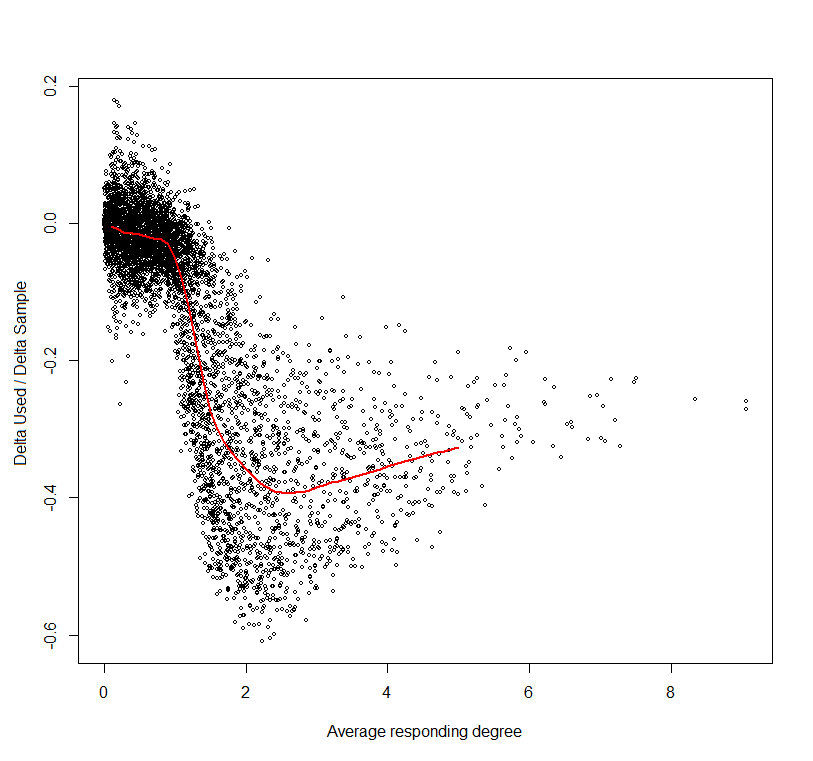}
\end{center}
\vspace{-8.0mm}
\caption{$\Delta used / \Delta Sample$ statistic, as a function of the population average degree among 2500 generated networks in a simulation run}
\end{figure}

Three other sampling order-based statistics are used in this study - all but one computed from the slope of a linear model of some node information.  The statistic $\Delta degree / \Delta sample$ is the rate that the average degree of sampled nodes changes throughout a sample. For a range of edge density, $\Delta degree / \Delta sample$ is likely to be negative because nodes with high degree are more likely to be reached early in a sample.

In the simulation study in Section 5, an infection mechanic is introduced to show the utility of this network inference method. Consequently, the statistic $\Delta infect / \Delta sample$ is used. This statistic represents the difference in the proportion of nodes that are infected from the beginning to the end of the sample.

For more work on incorporating sampling order information into a sample using network structure see (Crawford 2014)\nocite{crawford2014graphical} and (Crawford, Wu, \& Heimer 2015).

\section{INFERENCE WITH LINK-TRACED SAMPLES USING APPROXIMATE BAYESIAN COMPUTATION}

We are interested in inferring useful parameters of the network, such as degree distribution, rather than the specific structure of the network. We estimate these parameters with a strategy that involves a prior specification step, iterations of population generation steps and sampling steps, and a kernel density estimation step.

\subsection{Overview}

From a sample of interest, we compute a set of network statistics $s$.  We have a set of parameters about the population whose values we wish to infer from the observed sample. For those parameters, we define a joint prior distribution with support $\Omega$. 

For each of many simulations, we randomly generate a parameter set $p^*$ from the specified prior, and generate a population according to that parameter set. From that population, and possibly further guided by the parameter set $p^*$, we take a sample of the population by link-tracing. We compute summary statistics $s^*$ from the sample.

After the simulations are completed and we have a collection of parameter sets $p^*$ and their consummate statistic sets $s^*$, we employ kernel density estimation over the Cartesian product of $\Omega$ and $\mathcal{S}$, where $\mathcal{S}$ is the space of possible statistic sets. From that density function, we factor out the prior, and condition on the observed statistics s to obtain a model of the conditional density across $\Omega$.

\subsection{Prior specification step}

Parameters used in the simulation could determine the links per node, the response rates of recruits in general, or the response rates of important subsets of recruits. Other parameters determine the level of preferential attachment in link selection, or determine the propensity of nodes of form links with those similar to themselves. Similarity is defined here by relative proximity in some social or physical space, much like the network social space being inferred in Hoff et al. (2002). \nocite{hoff2002latent}

These parameters are to be estimates from three components:

A set of statistics that describe the sample of samples being observed, s. Let the set of statistics be s and let $\mathcal{S}$ be the space of possible s values.

Let $\Omega$ be the Cartesian product of the assumed ranges \\ $(Pmin_1 , \ldots , Pmin_{NP}$ to $Pmax_1 , \ldots , Pmax_{NP}$ respectively) of the parameters. 

A set of points in the Cartesian product of $\mathcal{S}$ and $\Omega$, $<\mathcal{S},\Omega>$ where points $p^*$ in $\Omega$ are selected from a joint random probability distribution, and s* is the result of a link-traced sample from a population described by $p^*$.

Possible probability distributions for the values of $p^*$ include uniform, discrete uniform, and geometric and beta with location-scale transformations. Using non-uniform distributions requires a weighting adjustment, as described in the kernel density estimation step in Section 4.5.

Specification of the priors is a balancing task between result reliability and computational efficiency. If the global maximum of the density is outside of $\Omega$ or simply at the periphery of $\Omega$, then it may be missed. To avoid this, boundaries for $\Omega$ that include all plausible areas, such as $0 < p < 1$ for response-to-recruitment probability, and $0 < p < 8$ for average link density, may be used.

For each simulation, a parameter set $p^*$ is taken with probability according to the prior distribution and within $\Omega$, and a population is randomly generated, according to $p^*$. However, if $\Omega$ is very inclusive, then many generated parameter sets will lead to samples with statistics far removed from the observed statistics, and thus will be of little use in estimating the true parameter values. This problem can be mitigated by generating many populations. Section 6 proposes some strategies for improving the selection of $\Omega$.

\subsection{Population generation step}

To generate a population, first generate the set of parameter values $p^*$ according to the given prior, including $N_V$ and $N_E$, the number of nodes and edges in this population respectively. For each node, randomly assign pertinent characteristics according to $p^*$, such as the location of the node in the unit square of space, link propensity, and any application-specific variables that are desired.

To form an edge, select a node $v_i$, either by simple random sampling, or by a weighted sample if a distribution of edge propensity is selected. Given $v_i$, select node $v_j$ with probabilities proportional to $\mathrm{D}^{- \gamma}$, where $D$ is a distance measure such as Euclidean distance in physical or social space, and $\gamma$ is a parameter in $p^*$ defining a level of tendency for nodes to form links with spatially close neighbours. The assignment of edges between nodes may be subject to application-specific variables, such as sex, sexual preference, the number of edges already attached to nodes, and dynamic markers like infection state.


\subsection{Link-tracing step}

We are given $n_{samp}$, the number of nodes to be sampled from the population by link-tracing. These are selected from the population according the following algorithm:

\begin{algorithm}[H]
 
 Step 1. If this is the first node, or the sampling queue has been exhausted, select a node by simple random sampling from those not yet selected. (After the first node this is known as a leap). Add the selected node, $X$, to the sample and skip to Step 4.\\
 Step 2. If there are nodes in the queue, remove any nodes that already included in the sample. For each node $X$ that is removed from the queue for already being in the sample, look up $source(X)$, which is the node that led to $X$, and increment the count of redundant links for $source(X)$. \\
 Step 3. If there are no remaining nodes in the queue, return to Step 1. \\
 Step 4. Query for the list of nodes in the population that are connected to $X$. For each connected node $sink(X)$, mark the node as responding with probability Pr(response) determined from $p^*$. Increment the number of links reported for node $X$. \\
 Step 5. Append the nodes marked as responding to the end of the sampling queue, and increment the number of links responding to $X$ for each such node. If the number of nodes in the sample is now equal to $n_samp$ or the number of the nodes in the population, stop. Else, return to Step 1. \\
 \caption{Example protocol used for sampling nodes by link-tracing}
 
\end{algorithm}

\subsection{Kernel density estimation step}

The values for given parameter for the set of simulations is i.i.d. realizations of some specified prior density $f(p)$.  Each parameter is generated independently, so the joint prior distribution is the product of individual priors, $f(\bar{p}) = \prod_i^{NP} f(p_i)$, where $NP$ is the number of parameters.

Each simulation produces a link-traced sample of its population. Statistics $s_i,\ldots$ for each sample are taken, producing an $Nruns \times NS$ matrix.

For each simulation, we have $NP$ and $NS$ parameter and statistic values, respectively. These values are collectively interpreted as a point  in $\Re^{(NP + NS)}$ space.  The $N_{runs}$ spatial points are scaled linearly to fit into a unit hypercube $[0,1] ^{(NP + NS)}$. Specifically, let $x_i^*$ be the $i^{th}$ unscaled parameter or statistic value of $NP$ and $NS$ such values respectively, and let $x_i$ be the scaled value.  For $i = 1 , \ldots , NP$, use the assumed parameter value bounds to scale, such that 


\begin{eqnarray}
x_i = (x_i^* - Pmin_i ) / (Pmax_i - Pmin_i). 
\end{eqnarray}

For $i = NP+1 , \ldots , NP+NS$, use the uniform method-of-moments estimates to scale, that is 

\begin{eqnarray}
x_i = (x_i^* - Qmin_i ) / (Qmax_i - Qmin_i), 
\end{eqnarray}
where $Qmin_i = ( min(x_i^*) - (1/(Nruns+1)) * (max(x_i) - min(x_i))$, and $Qmax_i = ( max(x_i^*) + (1/(Nruns+1)) * (max(x_i) - min(x_i))$.

After rescaling, a posterior density is estimated by kernel-smoothing Nruns point masses with an independent multivariate Gaussian-kernel.

Posterior density $(x_1 , \ldots , x_{NP}, x_{NP+1}, \ldots , x_{NP+NS}) =$

\begin{eqnarray}
G(x,y) =  (2\pi)^{-(NP+NS)/2}  \sum_{i=1}^{N_{runs}}{  \exp \left( \frac{ \sum_{j=1}^{N_p + N_s}( y_j - x_{ij})^2 ) }{ det(\Sigma)} \right) },
\end{eqnarray}

where $det(\Sigma)$ is the determinant of the variance-covariance matrix of the multivariate-normal, which is $\prod_j^{NP+NS}{\sigma_j}$ by the independence constraint placed on the kernel. However, since it is the likelihood, not the posterior, of ultimate interest, consider the result of kernel-smoothing points with masses inversely proportional to their prior densities:

\begin{eqnarray}
H(x,y) =  (2\pi)^{-(NP+NS)/2}  \sum_{i=1}^{N_{runs}}{ w_i \exp \left( \frac{ \sum_{j=1}^{N_p + N_s}( y_j - x_{ij})^2 ) }{ det(\Sigma)} \right) },
\end{eqnarray}
where $w_i =  1 / f(\bar{p})$.

Finally, to obtain the likelihood information $h(X|y)$, we condition $H(X,Y)$ on y*, the vector of statistic values from the actual sample,

\begin{eqnarray}
h(X | Y \approx y^*) = h(x | y^*) = \frac{ H(x,y^*)}{ \int_{x \in X} { H(x,y^*) dx}} .
\end{eqnarray}
$h(X|y^*)$ is not a true conditional in that it incorporates information from all values of $y$. However, controlling for prior density, point masses with $y$ values near $y^*$ weigh more heavily on the modeled distribution of $X$. That is, the contribution of a simulated sample increases with its proximity of its statistics to $y^*$. To plot the probability density over a range in order to observe the posterior maximum, mean, and the credible interval spatial smoothing methods such as kriging are employed.

\section{Simulation study}

\subsection{Overview}

Consider the network graph shown in the two panels of Figure 3. This is a network of order 400 taken from a population of unknown size by a link-tracing design. Each point represents a node, and each line segment represents a undirected link. Both panels show the same sample. The left panel includes the links that were reported, but not used for recruitment. In the right panel, only links that led to a recruitment into the sample are included. Notice that in the left panel, some of the only have a node at one end; the missing nodes were not recruited and therefore not observed directly, so they do not appear in the figure. 

The location of the points in Figure 3 represent the locations of each node (e.g. a person, animal, or computer) in a social or physical space. The shade and shape of each point represents an infection status; filled squares represent infected nodes and empty circles represent uninfected nodes. Many infection mechanics are possible, but for this study, we assume that a proportion of nodes have been infected by an outside source, and that this happened before any of the connections between nodes were formed. Let this initial infection proportion be $\phi$. After the initial infection step, we establish connections in a random order. For each connection made, a node $v_i$ is selected with equal probability to any other node, and node $v_i$ makes a connection with another node with probability proportional to $D^{-\gamma}$, where $D$ is the Euclidean distance between $v_i$ and the other node.

\begin{figure}[ht]
\begin{center}
\includegraphics[angle=0,scale=0.25]{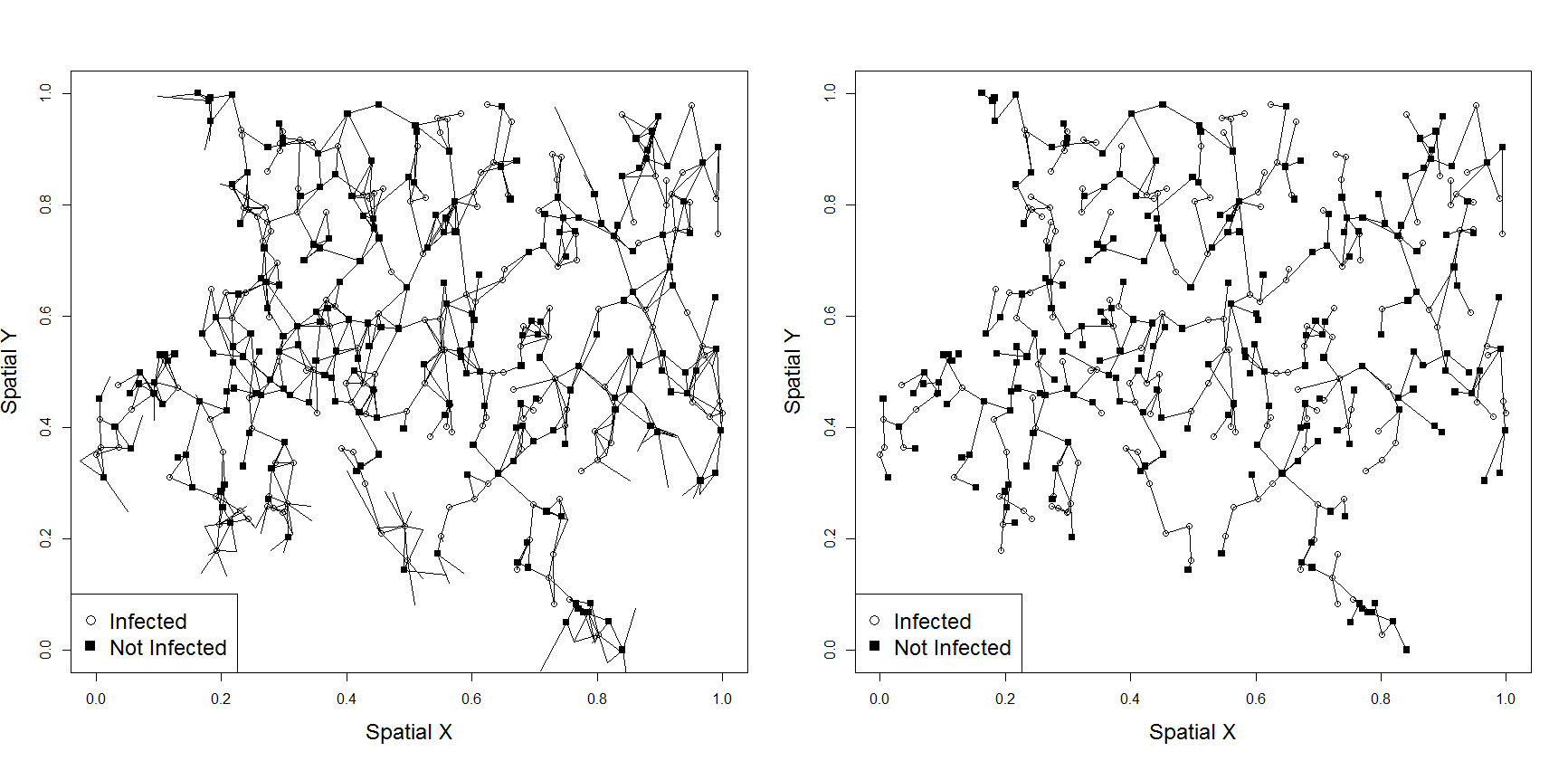}
\end{center}
\vspace{-8.0mm}
\caption{ Sample of order 400 from the population of interest, with all the edges found in link-tracing in the left panel, and with only the recruiting edges retained in the right panel}
\end{figure}

We wish to infer five parameters: \\
1) The average degree of the population, \\
2) the number of nodes in the population, $N_V$ \\
3) the proportion of nodes that initially have the `infected' status, $\phi$, \\
4) the transmission chance, $\alpha$, and\\
5) the `nearness preference' parameter, $\gamma$. 

We will assume for simplicity that selection into the sample, by either simple random selection or by link-tracing, never fails. In short a perfect response rate is assumed. We also assume the only factor determining edge formation is distance, and that there are no latent effects such as unequal edge propensity or preferential attachment based on degree or other node variable (e.g sex, social status, known infected status).

Recall from Section 4 that to use the proposed method, a prior and a set of summary statistics are required. Having neither prior nor statistics given, we conduct two rounds of ABC Network Inference: one to determine a reasonable prior and set of statistics, and a more computationally expensive round using said prior and statistics.

In the initial round, we simulate 500 populations using parameter sets from a non-informative joint prior distribution. This initial set is used to inform the selection of statistics to condition on to obtain a function for the conditional probability density $h(X|y*)$. We then use kernel density estimation to compute the conditional density throughout points in the parameter support $\Omega$ to decide if the prior needs to be changed to better find the global maximum, or if a large portion of the support isn't simulating over.

In the final round, we proceed similarly, but simulate 2500 populations using an improved prior, and condition the posterior from the KDE on the statistics identified from the first round rather than going through a statistic selection process.


\subsection{Initial Round}

For the five parameters of interest, we use a joint prior of the product of population average degree $\sim unif(0,7)$, $N_V \sim 200 + geom(mean=1000)$, $\alpha \sim unif(0,0.5)$, $\phi \sim unif(0,0.3)$, and $\gamma \sim unif(-2,10)$.

Available are the following summary statistics from the sample: \\
1) Mean degree recruited $ = 1.995$\\
2) mean degree reported (among all nodes) $ = 3.908$\\
3) mean degree reported (among only infected nodes) $ = 4.104$\\
4) mean degree recruited difference (infected $-$ not infected) $ = 3.999$ \\
5) sample infection proportion $ = 0.560$\\
6) $\Delta degree / \Delta sample = -0.0798$ \\
7) $\Delta depth / \Delta sample = 0.602$ \\
8) $\Delta used / \Delta sample = -0.139$ \\
9) $\Delta infect / \Delta sample = -0.086$ \\

The statistic $\Delta degree/ \Delta sample$ is defined as the average rate of change in reported degree among sample units from the beginning of sampling to the end of sampling. As a simplified example, if the first few units sampled had mean degree 5, and the mean degree decreased linearly as the network were explored until the mean degree was 2 at the end of the sample, then  $\Delta degree/ \Delta sample \approx -3$ for that sample, 

The statistic $\Delta depth / \Delta sample$ is defined as the average rate of change of depth from the beginning to the end of sampling. The depth of a node, in this instance, means the geodesic distance to any other connected node, averaged over those connected nodes. Well-connected and central nodes have low depth, and nodes on the periphery of a network have high depth. The hypothesis behind using this statistic was that in networks where the sample was a large proportion of the population that the low-depth nodes would be exhaused before the sample had completed, and that the averge depth of the nodes being sampled would increase.

The statistics $\Delta used / \Delta sample$, and $\Delta infect / \Delta sample$ are computed similarly. The term `used' refers to the proportion of a node's connections that are used for recruitment. The term `infect' refers to a binary variable of infected status. In the case of  $\Delta infect / \Delta sample$, the log-odds ratio was used, instead of the linear regression slope.

We take a link-traced sample from each of 500 simulated populations guided by the parameter sets from this prior. We compute all the candidate summary statistics from each sample. For each of the five parameters Y and the each of the nine candidate statistics $X$, we fit a cubic regression, $Y \approx \beta_0 + \beta_1X + \beta_2X^2 + \beta_3X^3$. Statistics that are strongly related to the parameters of interest are likely to fit a cubic regression better than a null model. The coefficients of determination, $R^2$, and the F-statistics for each model comparison are shown in Tables 1 and 2 respectively.

\begin{table}[ht]
\centering
\begin{tabular}{rrrrrr}
  \hline
Sample Statistic & Avg.Degree & NV & Init. Infection & Pr(Infect) & Closeness Param \\ 
  \hline
  Mean Deg. Recruited & 173.02 & 3.61 & 8.54 & 1.11 & 1.61 \\ 
  Mean Deg. Reported & 6051.57 & 1.40 & 11.37 & 0.51 & 1.84 \\ 
  Mean Deg. Infected & 4794.38 & 1.22 & 11.80 & 1.13 & 1.86 \\ 
  Mean Deg. Difference & 22.87 & 1.27 & 12.27 & 13.89 & 3.00 \\ 
  Infection Prop. & 16.31 & 0.27 & 11674.92 & 17.38 & 0.40 \\ 
$\Delta degree / \Delta sample$ & 3.82 & 39.46 & 1.29 & 0.46 & 13.61 \\ 
$\Delta depth / \Delta sample$ & 48.68 & 11.79 & 3.65 & 0.76 & 24.45 \\ 
$\Delta used / \Delta sample$ & 186.58 & 8.28 & 9.67 & 1.00 & 120.54 \\ 
$\Delta infect / \Delta sample$ & 2.91 & 2.05 & 9.77 & 3.18 & 2.50 \\ 
   \hline
\end{tabular}
\caption{F-statistics of models of parameters of interest by summary statistics}
\end{table}

\begin{table}[ht]
\centering
\begin{tabular}{rrrrrr}
  \hline
Sample Statistic & Avg.Degree & NV & Init. Infection & Pr(Infect) & Closeness Param \\ 
  \hline
Mean Deg. Recruited  & 0.58 & 0.03 & 0.06 & 0.01 & 0.01 \\ 
Mean Deg. Reported & 0.98 & 0.01 & 0.08 & 0.00 & 0.01 \\ 
Mean Deg. Infected & 0.98 & 0.01 & 0.09 & 0.01 & 0.01 \\ 
Mean Deg. Difference & 0.16 & 0.01 & 0.09 & 0.10 & 0.02 \\ 
Infection Prop. & 0.12 & 0.00 & 0.99 & 0.12 & 0.00 \\ 
$\Delta degree/ \Delta sample$ & 0.03 & 0.24 & 0.01 & 0.00 & 0.10 \\ 
$\Delta depth / \Delta sample$ & 0.28 & 0.09 & 0.03 & 0.01 & 0.16 \\ 
$\Delta used / \Delta sample$ & 0.60 & 0.06 & 0.07 & 0.01 & 0.49 \\ 
$\Delta infect / \Delta sample$ & 0.02 & 0.02 & 0.07 & 0.03 & 0.02 \\ 
   \hline
\end{tabular}
\caption{$R^2$ of models of parameters of interest by summary statistics}
\end{table}

Many different summary statistics can be used, but since each one increases the dimensionality of the kernel density estimation step, extraneous ones should be avoided.
 
 In this simulation, average population degree is very well predicted from the average reported degree of respondents. Assuming that we have the number of edges of each node (even if we don't know what the edges are), then estimating population average degree by approximate Bayesian computation is unnecessary. From the cubic regression model using sample average degree, the 99\% prediction interval for population average degree is 2.772 to 4.290. The initial prevalence of the infection is predicted well from the sample infection prevalence, but still has some variation left to explain.

 The parameter of closeness preference, $\gamma$, is partially predicted by $ \Delta used / \Delta sample$. Figure 4 is a scatterplot of values from the 500 simulations showing the interaction between $\gamma$, $ \Delta used / \Delta sample$, and the average degree of the population. For networks with an average of 2 or more edges attached to each node, there is an increasing drop-off in reported edges that lead to recruitment.  When the closeness preference is large ($ 6 \leq \gamma \leq 10$), the change in proportion of links used is small (0 $ \leq \Delta used / \Delta sample \leq $ 0.2).

\begin{figure}[ht]
\begin{center}
\includegraphics[angle=0,scale=0.40]{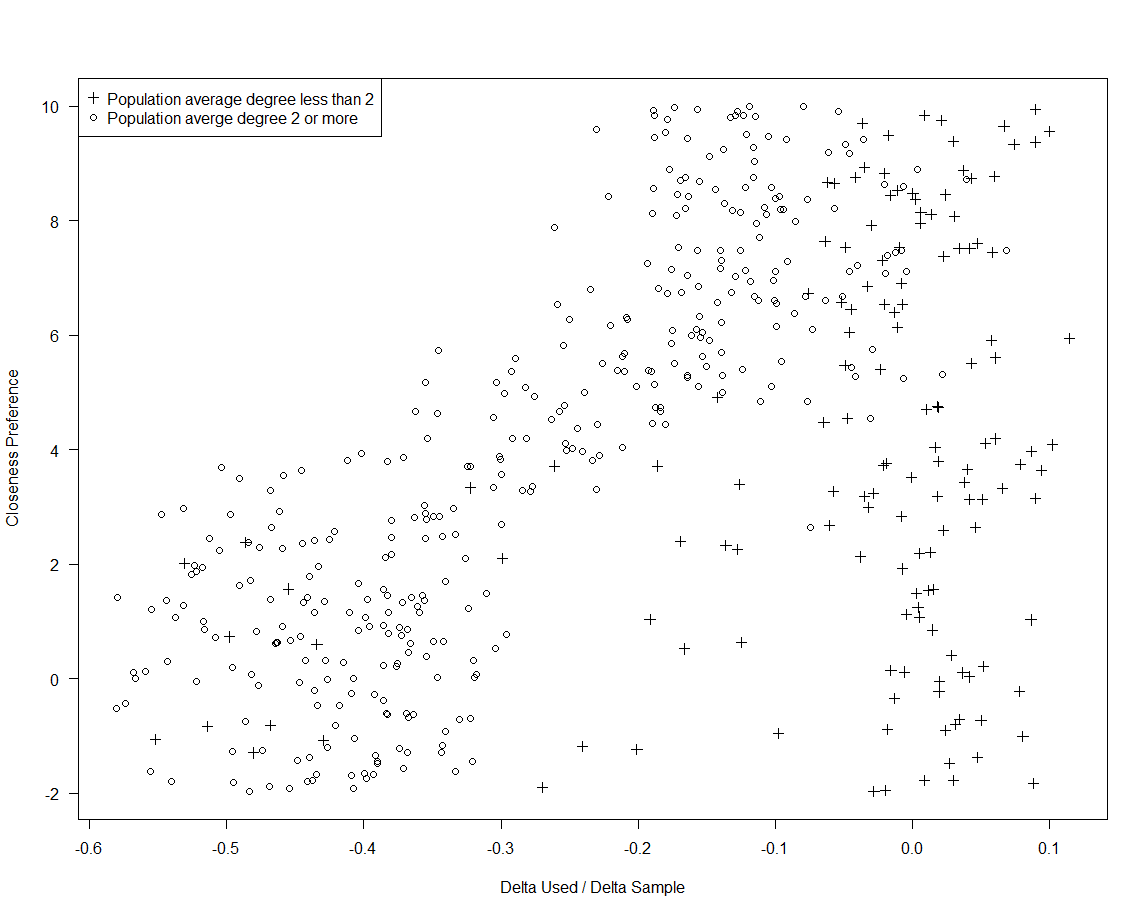}
\end{center}
\vspace{-8.0mm}
\caption{$\Delta used / \Delta sample$ statistic, closeness preference parameter $\gamma$ and population average degree among 500 randomly generated networks in an initial simulation run}
\end{figure}

 For networks with average degree between 1 and 2, the relationship is less pronounced. For networks with average degree less than 1, no relationship between $\gamma$ and  $ \Delta used / \Delta sample$ is apparent. This is unsurprising as the samples from these networks are much more likely than their denser counterparts to fully explore a network component and move on to another.

 Population size is only marginally predicted by one of the chosen summary statistics - the rate that node degree decreases through a sample, $\Delta degree / \Delta sample$. Figure 5  shows how $\Delta depth / \Delta sample$ decreases dramatically when the sample of the network includes all or most of the nodes in the population. The dashed line is at 400 nodes; any points below the line represent samples that include all the nodes in the population. Any interaction involving the average degree in the population is tenuous at best.

\begin{figure}[ht]
\begin{center}
\includegraphics[angle=0,scale=0.40]{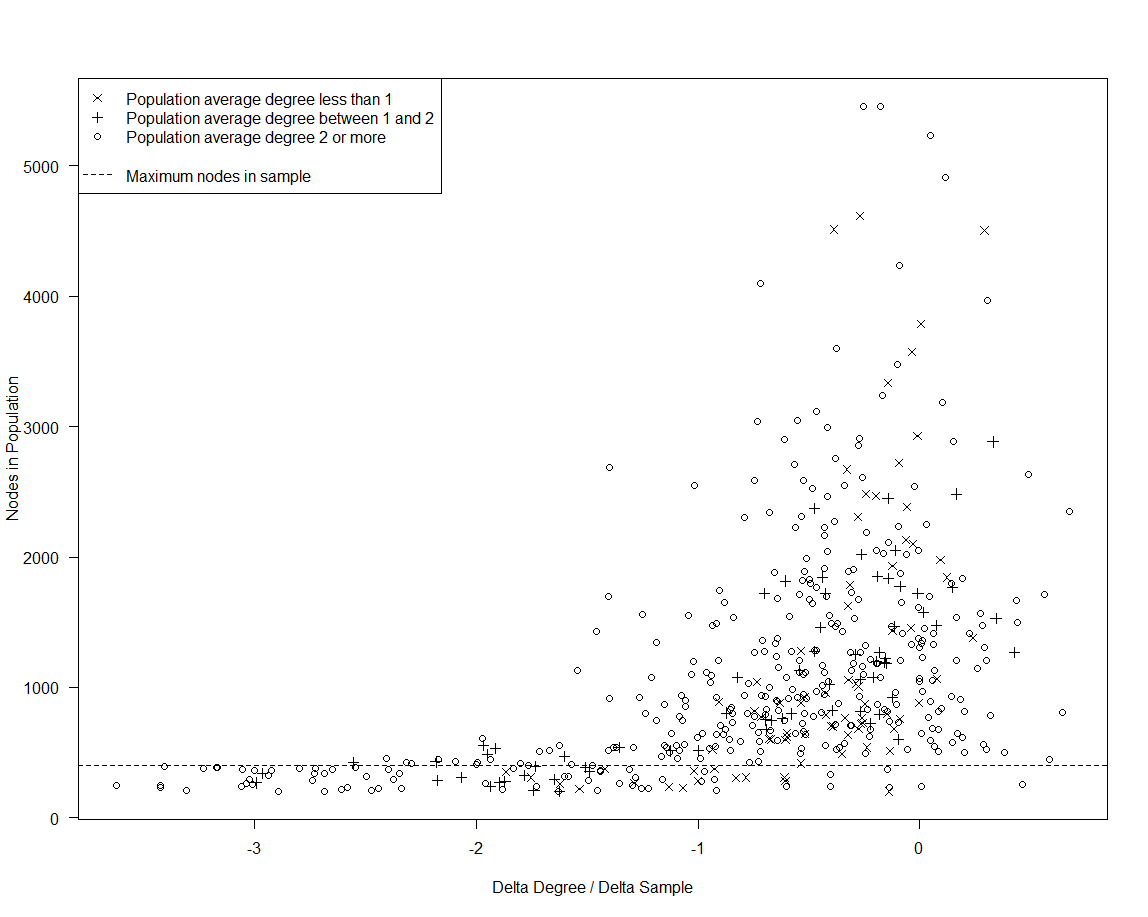}
\end{center}
\vspace{-8.0mm}
\caption{$\Delta Degree / \Delta sample$ statistic, population network order parameter $N_V$, and population average degree among 500 randomly generated networks in an initial simulation run}
\end{figure}

 None of the summary statistics show a substantial relationship with the probability of infection, $\phi$. The statistic with the strongest relationship is the observed infection proportion, which shows a relationship with $\phi$ in part as an artifact of the prior. The statistic with the second strongest relationship to infection probability, infection-degree differential, has no obvious pattern.

For the four parameters with no singularly determining factor, there are four potentially useful summary statistics: the infection-degree difference, the sample infection prevalence,  $ \Delta degree / \Delta sample$, and $ \Delta used / \Delta sample$; thus we have eight dimensions along which to do kernel density estimation. The six panels of Figure 6 show spatially-smoothed two-dimensional cross-sections of the results from the KDE step, conditioning on the observed summary statistics.  They show large regions of the space of population order and initial infection prevalence with almost no posterior density. Also, the boundaries of the support have a non-trivial share of the density. Informed by the initial kernel density estimation and linear predictions, the prior is updated to better include possible parameter values and to reduce the proportion of non-productive simulations.

\begin{figure}[htpb]
\centering
\mbox{\subfigure
{\includegraphics[width=3.4in]{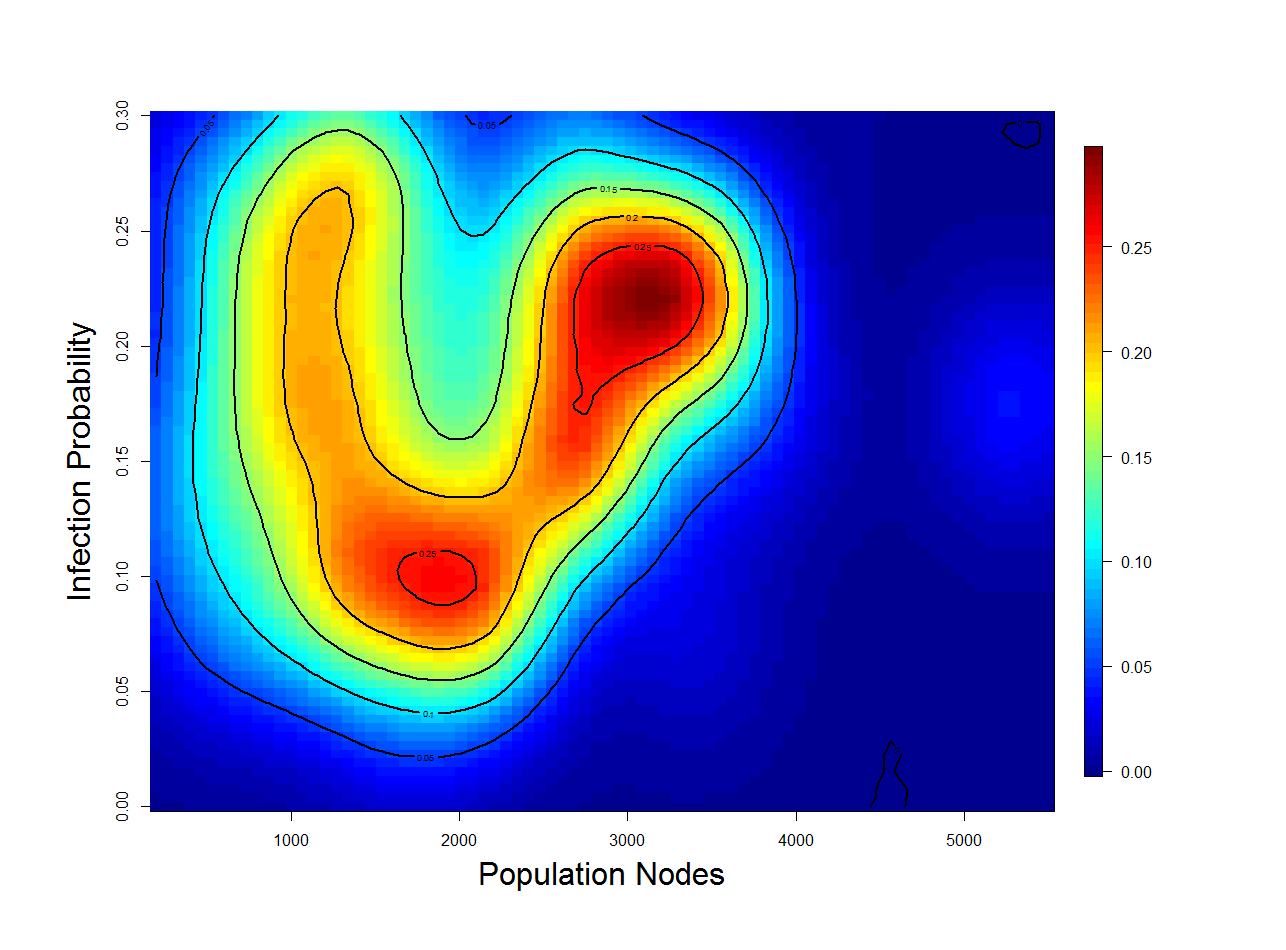}}
\subfigure
{\includegraphics[width=3.4in]{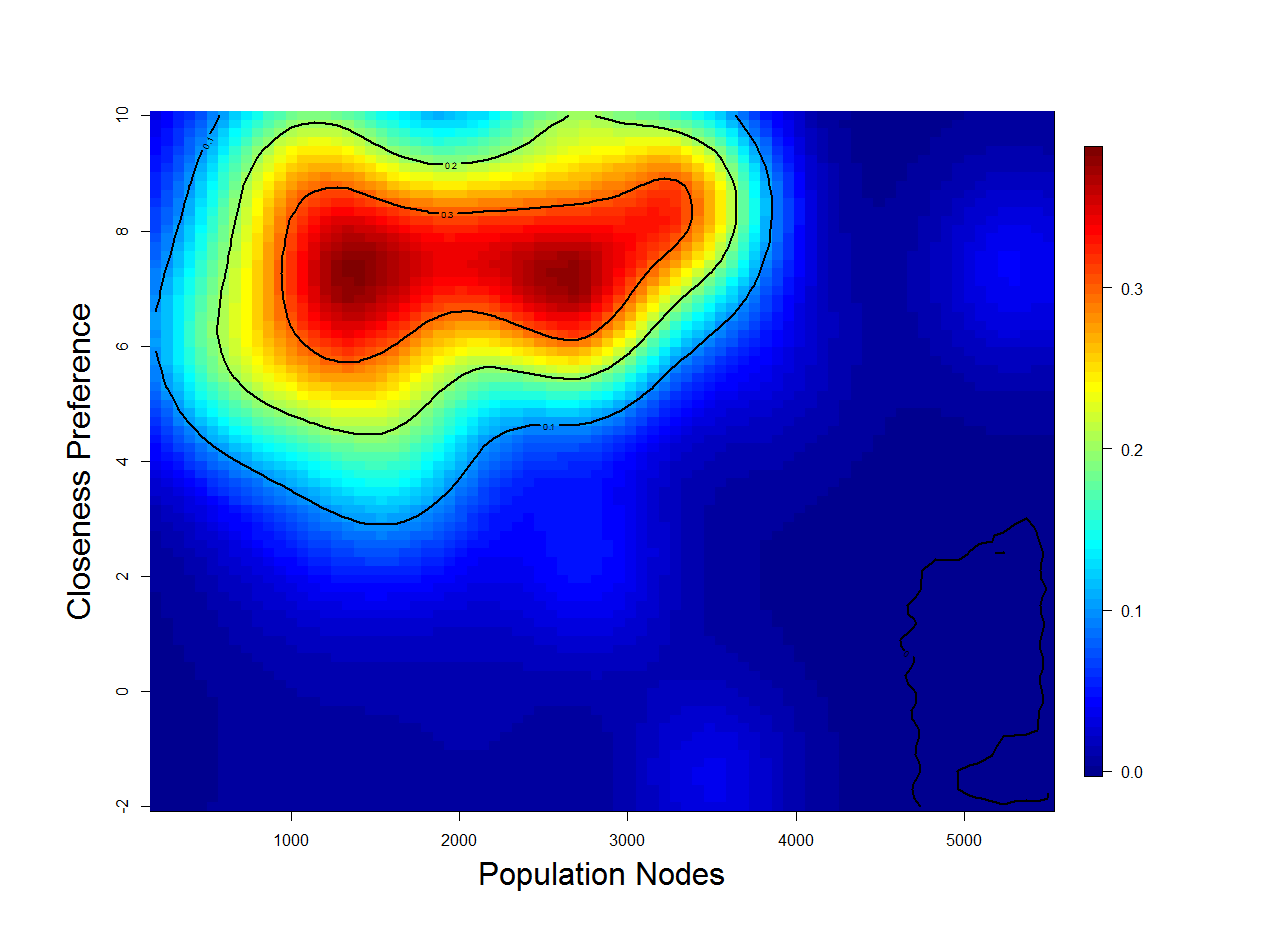}}}\\
\mbox{
\subfigure
{\includegraphics[width=3.4in]{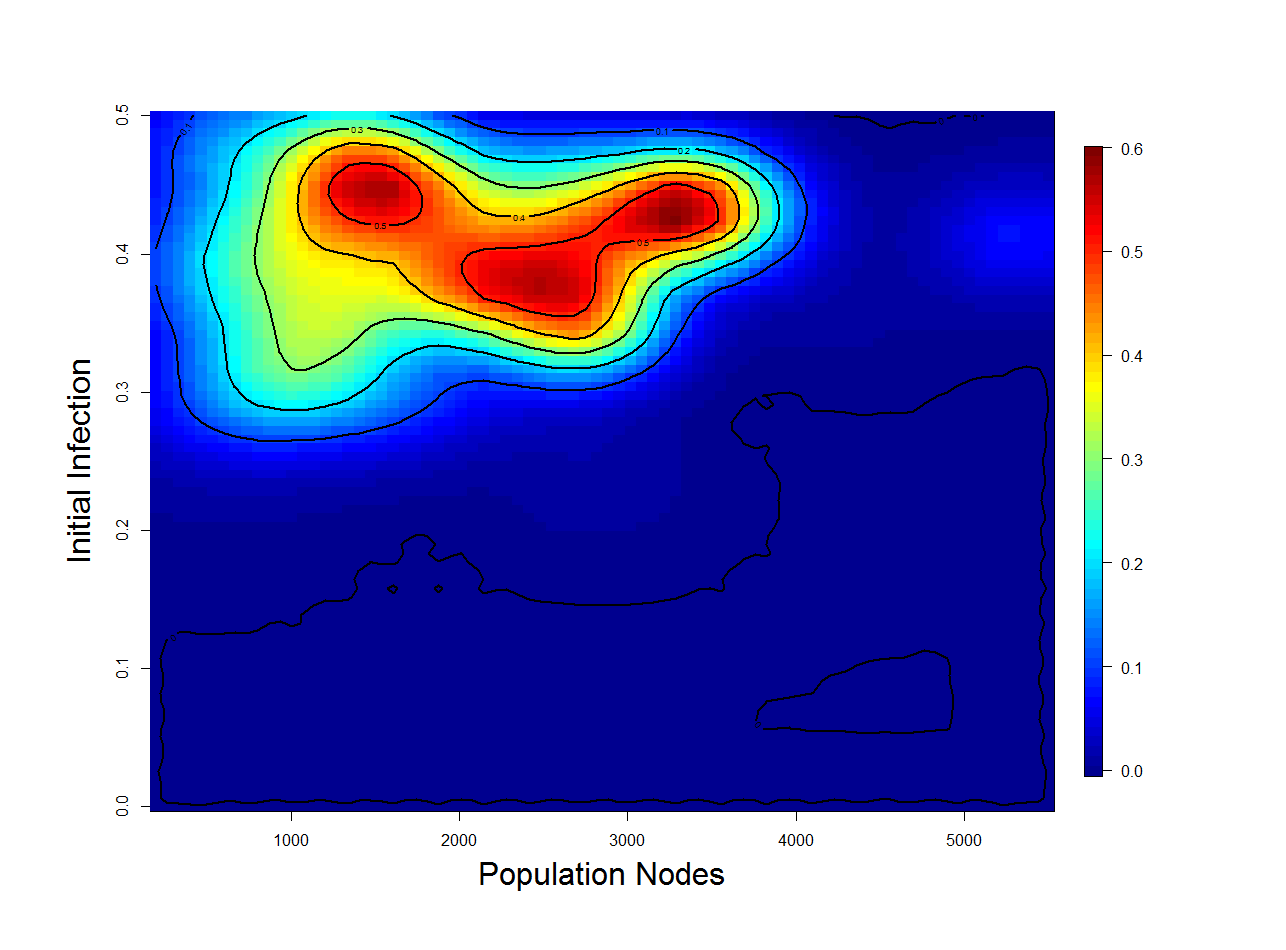}}
\subfigure
{\includegraphics[width=3.4in]{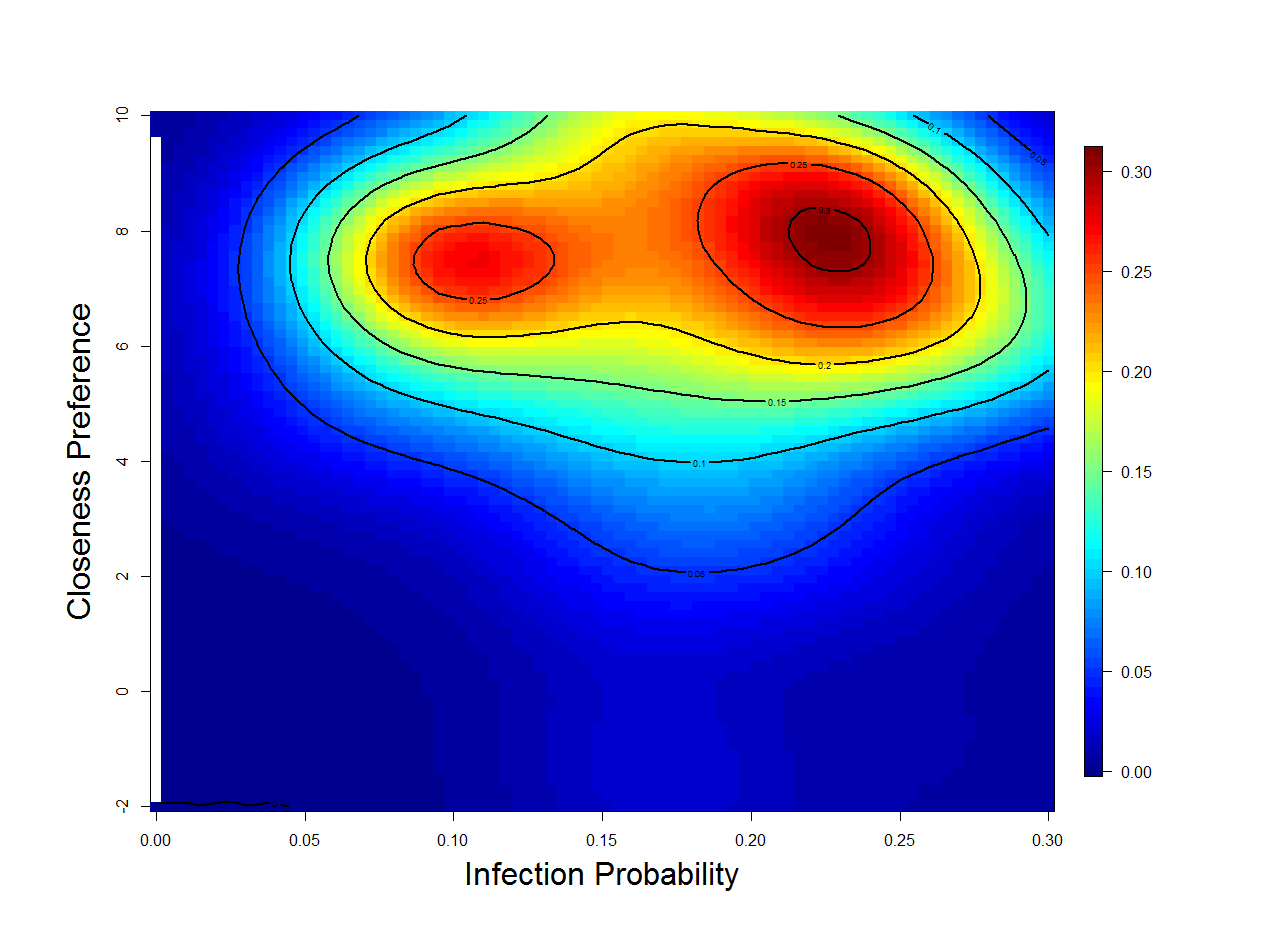}}}\\
\mbox{
\subfigure
{\includegraphics[width=3.4in]{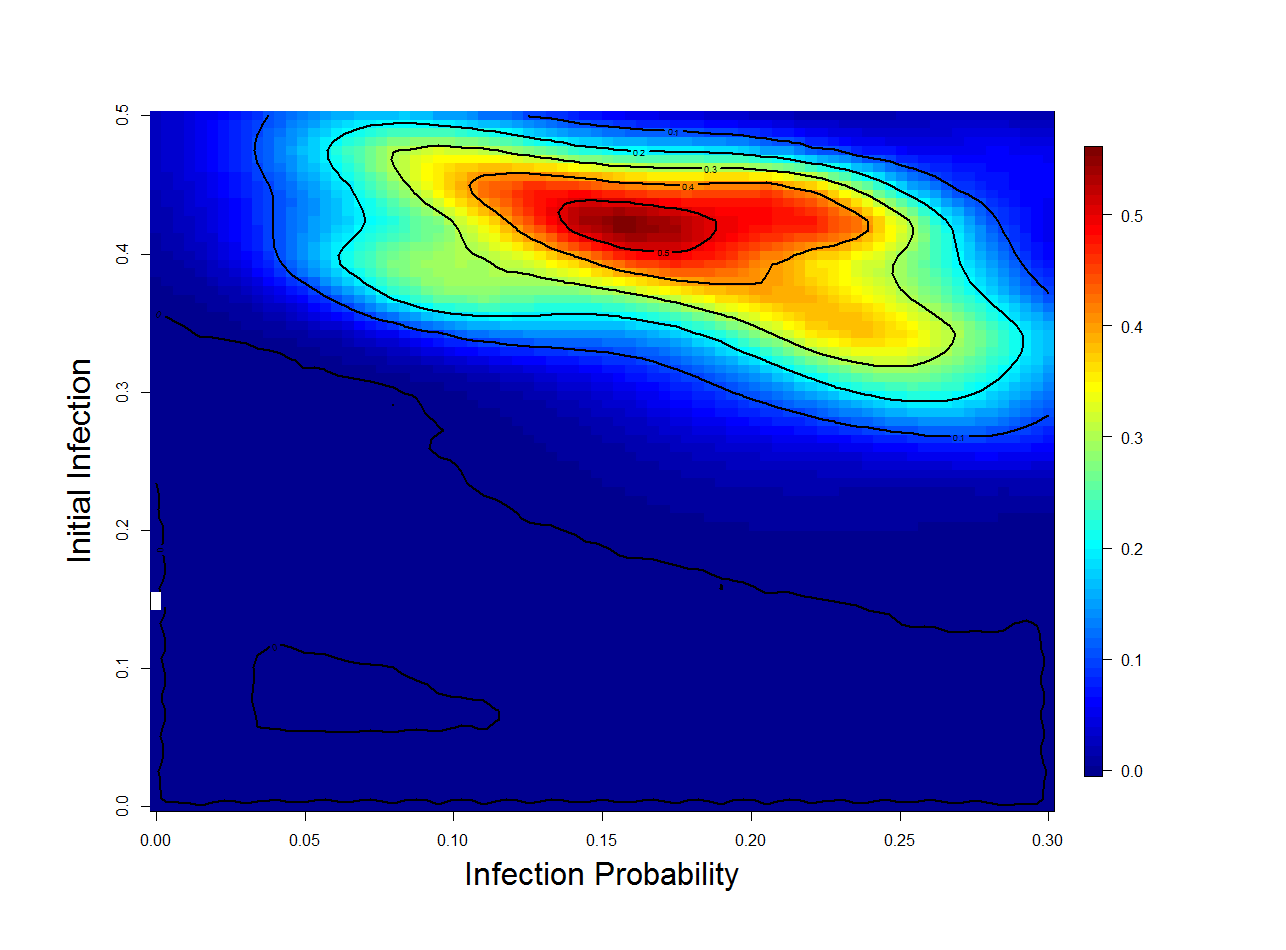}}
\subfigure
{\includegraphics[width=3.4in]{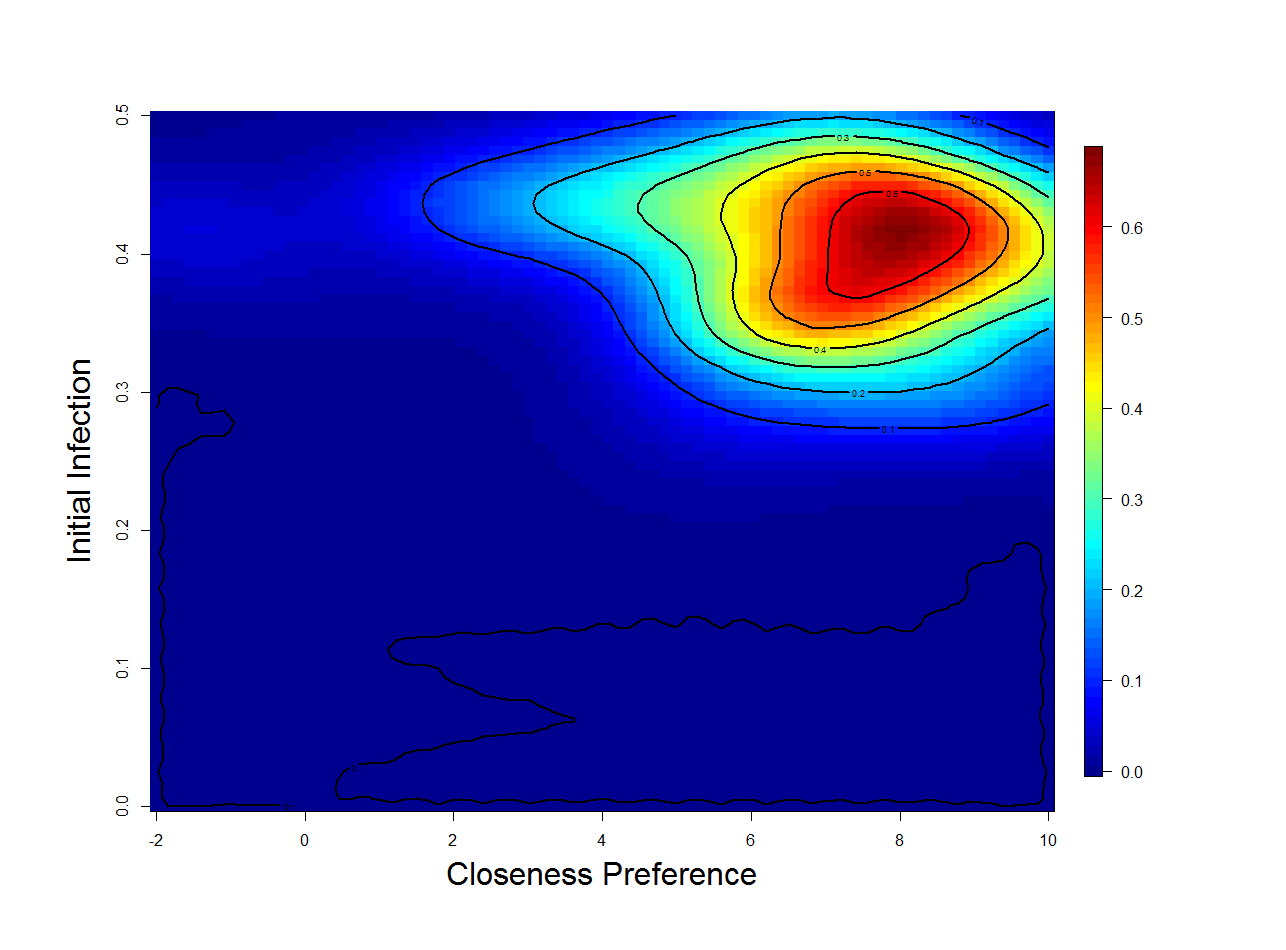}}}
\caption{ Initial round weighted-posterior density of infective probability, number of population nodes, initial infection proportion, and preference for closeness in forming connections, shown as two-dimensional marginals}
\end{figure}

\subsection{Final Round}

Using the information from the first round we now use a joint prior of the product population average degree $\sim unif(2.772,4.290)$, $N_V \sim 200 + geom(mean=1000)$, $\alpha \sim unif(0.2,0.6)$, $\phi \sim unif(0,0.4)$, and $\gamma \sim unif(2,12)$.

In this round take a link-traced sample from each of 2500 simulated populations, normalize, and use kernel density estimation on the joint space of $\Omega_1$, which is $\Omega$ with updated bounds a dimension for population average degree included, and $\mathcal{S}_1$, which is $\mathcal{S}$ with updated bounds. The panels in Figure 7 show the two-dimensional cross-sections of the conditional density from the improved estimates. The centroid of the conditional probability and the real parameter values are labelled as triangles and squares in each panel, respectively. The marginal posterior means, as well as the true values of the parameters are listed in Table 3. 


\begin{figure}[htpb]
\centering\mbox{\subfigure
{\includegraphics[width=3.4in]{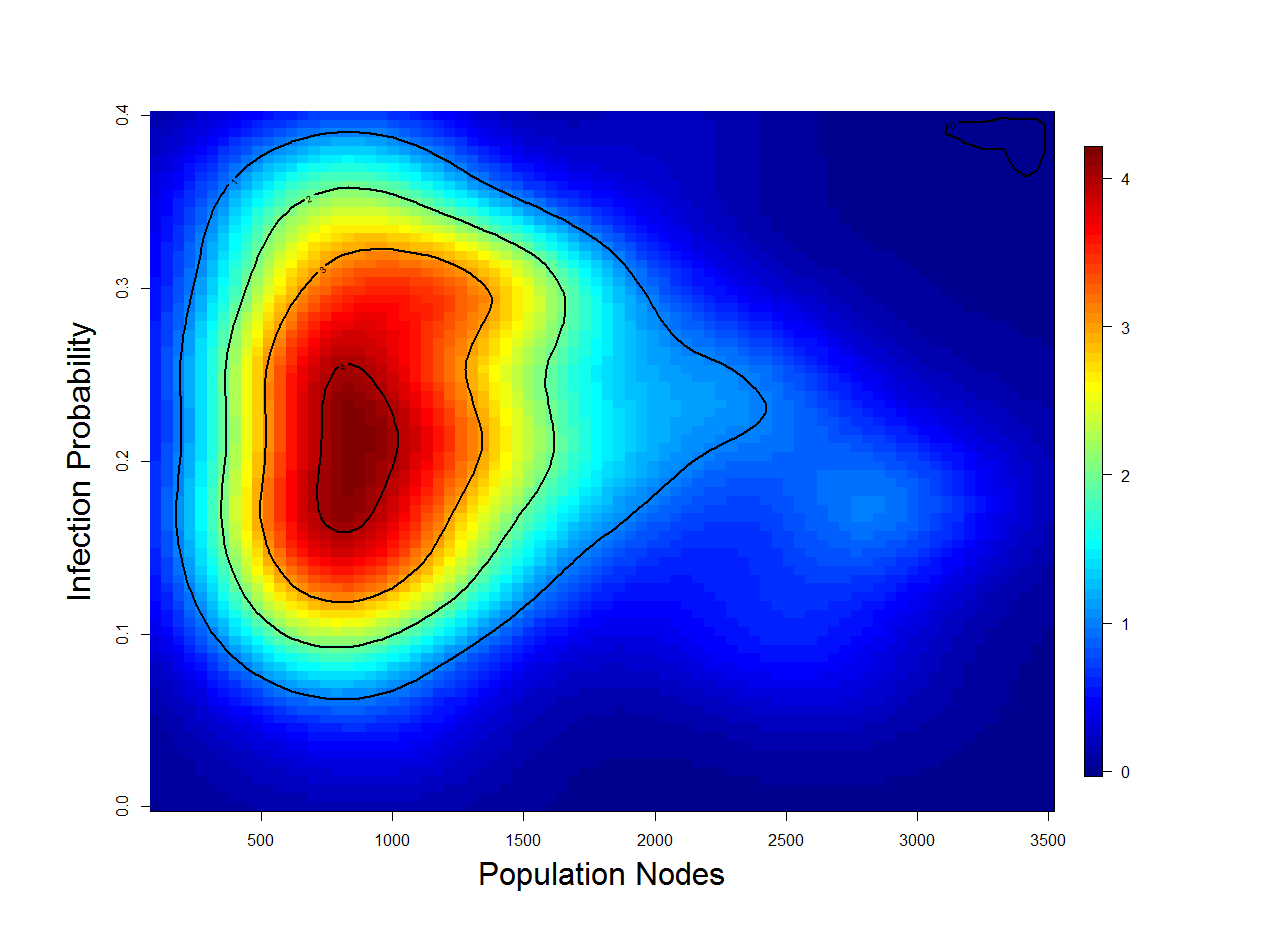}}
\subfigure
{\includegraphics[width=3.4in]{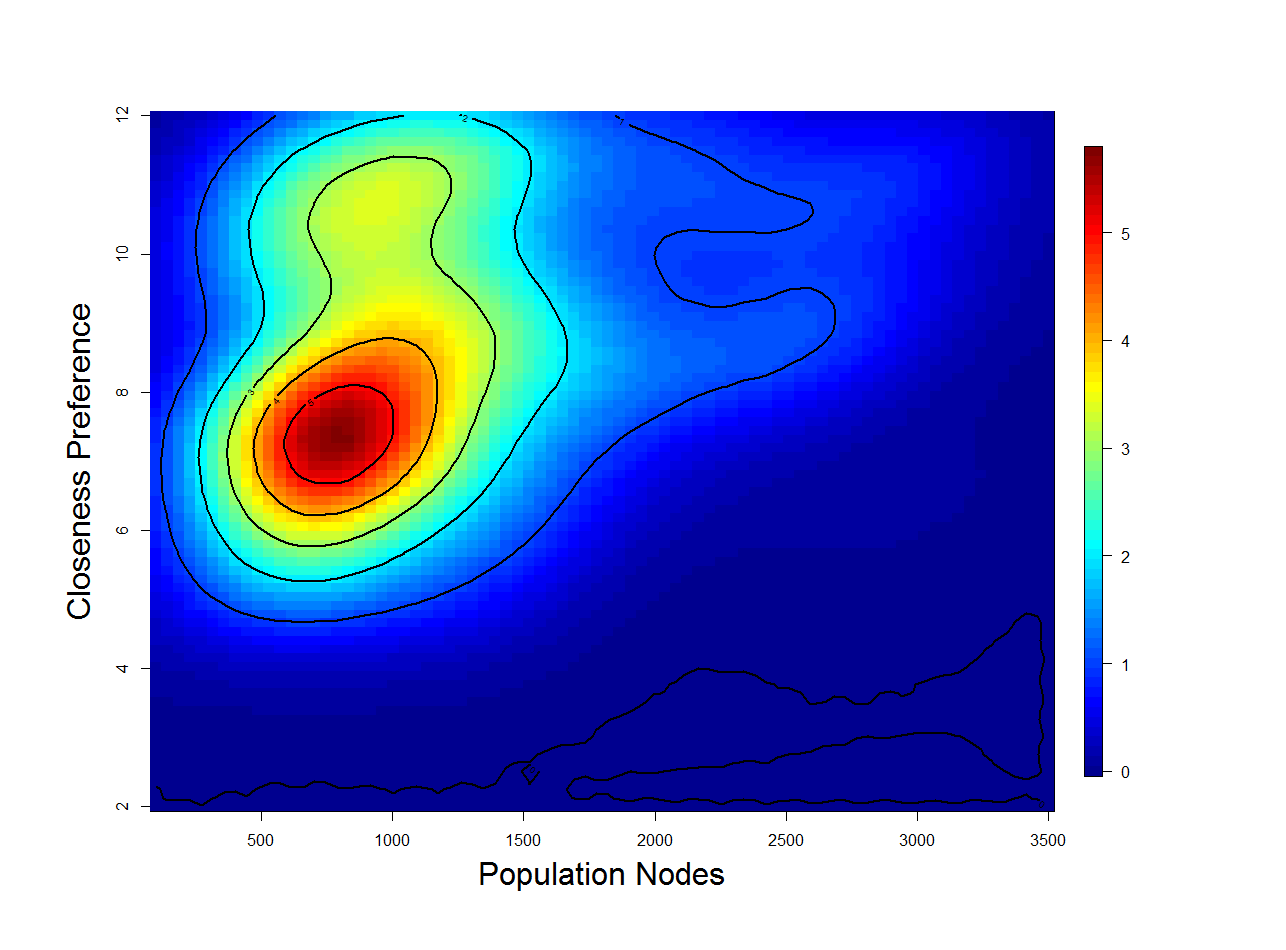}}}\\
\mbox{
\subfigure
{\includegraphics[width=3.4in]{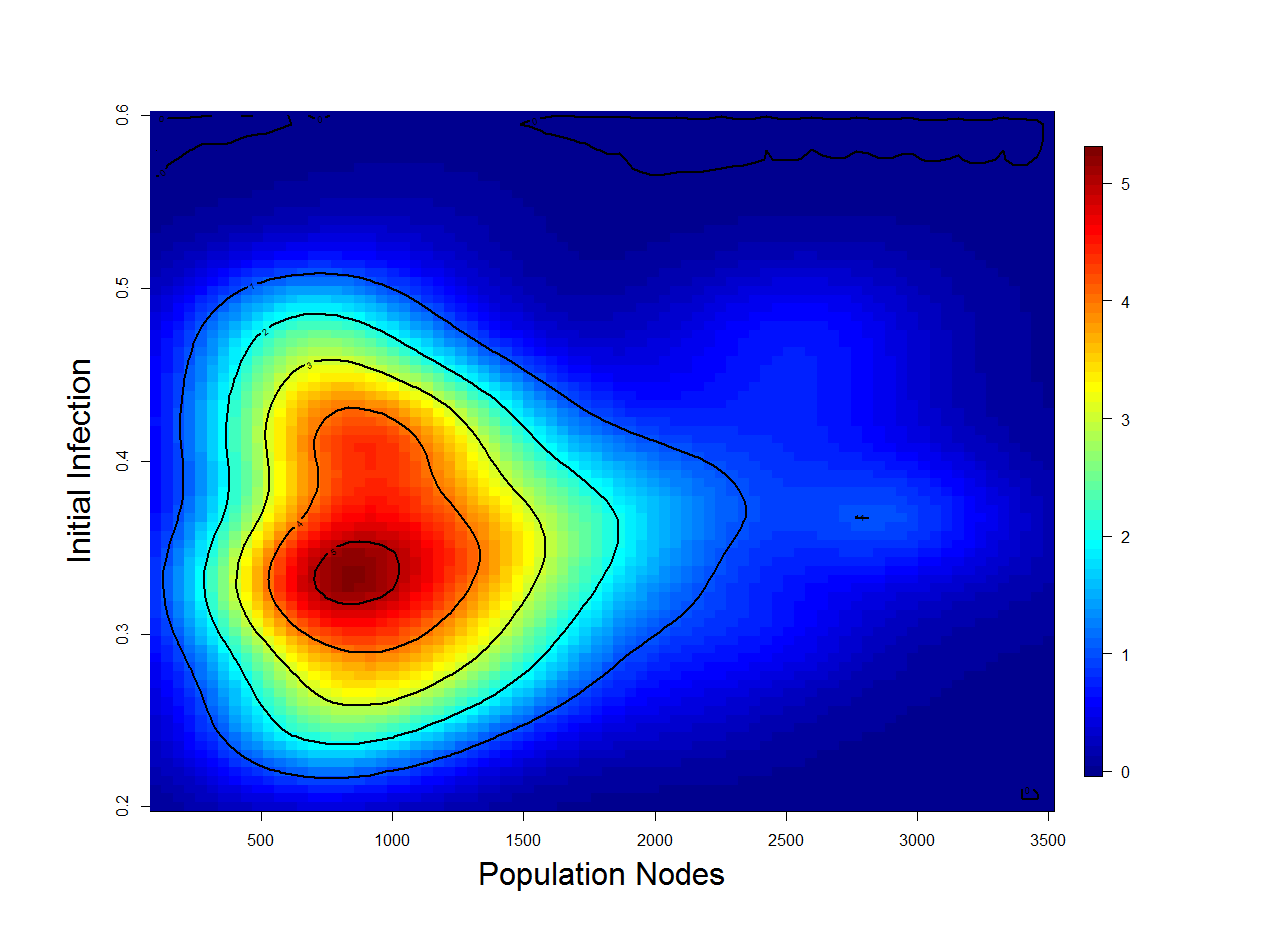}}
\subfigure
{\includegraphics[width=3.4in]{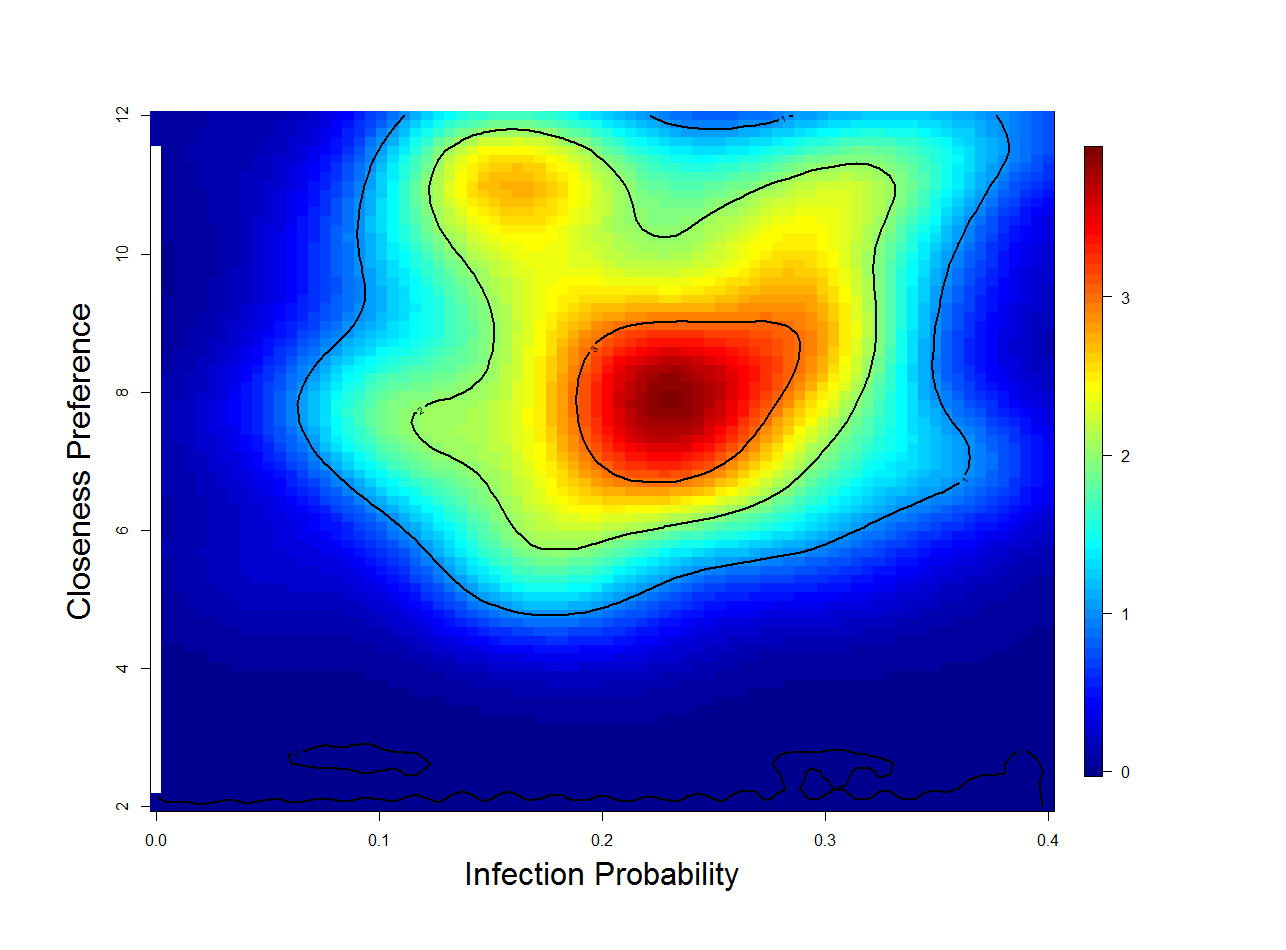}}}\\
\mbox{
\subfigure
{\includegraphics[width=3.4in]{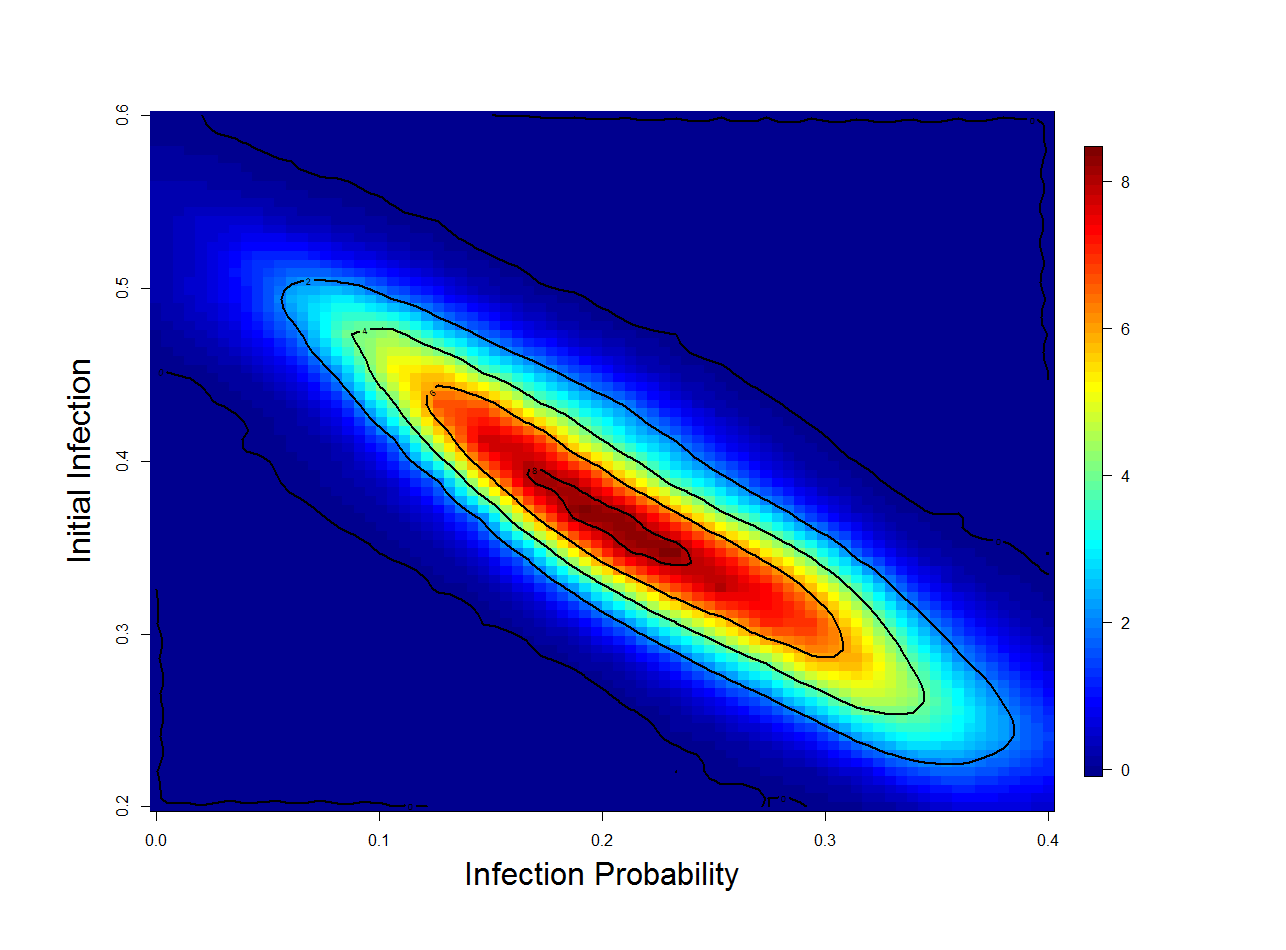}}
\subfigure
{\includegraphics[width=3.4in]{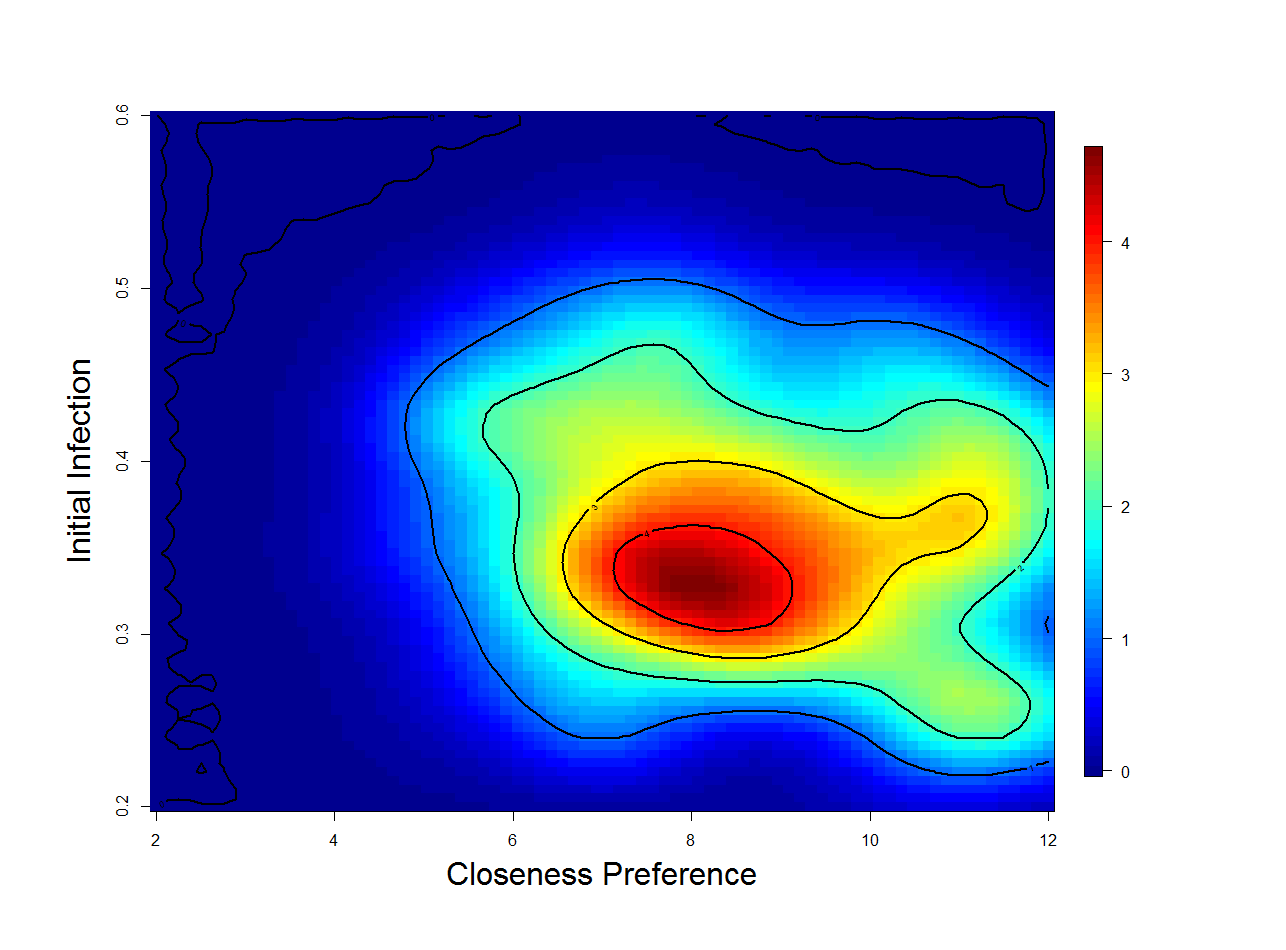}}}
\caption{ Final l round weighted-posterior density of infective probability, number of population nodes, initial infection proportion, and preference for closeness in forming connections, shown as two-dimensional marginals}
\end{figure}

\begin{table}[ht]
\centering
\begin{tabular}{lccc}
  \hline
Parameter & True & Mean (Initial Round) & Mean (Final Round)  \\ 
  \hline
 Number of Nodes 				& 1412  & 2049  & 2094   \\ 
 Initial Infection $\alpha$		& 0.467 & 0.399 & 0.374  \\ 
 Transmission Chance $\phi$		& 0.093 & 0.175 & 0.201  \\ 
 Nearness Preference $\gamma$	& 7.298 & 6.753 & 9.352  \\ 
   \hline
\end{tabular}
\caption{True parameter values and weighted-posterior mean after two simulation runs}
\end{table}

\section{Inference on the CanLII Database using Approximate Bayesian Computation}

In this section, we demonstrate the application potential of our to a naturally occurring, rather than a synthetic, dataset. The dataset examined in these sections is a collection of cases and citations of Canadian laws, centering around cases involving the Supreme Court of Canada (SCC). In many cases and other legal decisions, previous cases are cited as precedents to provide context to the situation being examined, and to ensure consistency. 

By treating each case as a node, and each citation as a directed edge from the citing case to the cited case, it becomes apparent that this collection is a network. This network is fully dynamic in time - new cases occurring represent a birth process of nodes, and the citations that new cases make to existing work represent a birth process of edges. We are interested in a small number of parameters governing which cases receive many citations and which receive few. Since the number of interesting parameters is small, and the dataset is a large, complex network, with complications such as a birth process, the extreme flexibility of Approximate Bayesian Computation makes it uniquely well suited to this problem.

We wish to do some preliminary work to identify 'dead' laws, which have not been used in a very long time. The Canadian Law Information Institute (CanLII) database contains more than 1 million publicly available documents of decisions from the Supreme Court of Canada and of each Canadian Province and Territory. It also includes includes similar documents from lower courts like the Federal Court of Canada, specific focus courts like (xx), and legislation and regulations from Parliament from each legislative assembly. Legislation is available as a set of aggregated documents, but old versions of the aggregate aren't available. 

The database includes all decisions from the Supreme Court of Canada's (SCC) decisions, which amounts to more than 10,600 documents over 140 years.  The database also includes all recent (i.e. the last 10-20 years) documents of lower courts, and some of the earlier documents. Years with incomplete coverage for a given court are identified in the database. For the sake of feasibility, our sample is the text of the SCC decisions, and the list of documents that cite each of these decisions.

The CanLII database represent a sample of the network of public legal documents in Canada. In this network, documents are represented by nodes and citations are represented as directed edges. In our sample of the network (the 10,600 Supreme Court decisions), each document only cites older documents and is cited by future documents, therefore the sample network is acyclic. This acyclic nature may not necessarily be the case for a wider range of documents, if multiple revisions of the document are considered a single node.

Most documents cite, and are cited by, other documents. For clarity, we refer to the previous documents that a given document cites as its 'parents', and any future documents that cite the document in question as its 'children'. If a document's parent appears in the CanLII database, a web hyperlink appears in the document to that parent. Likewise, if a document's child appears in the database, then a reference to that child appears in the document's 'cited by' list. A document may have parents or children that do not appear in the database. Out-of-database parents can be uniquely identified by scanning the text of a document for citations, which allows us to know the number of parents a document has and makes identification of co-citations possible, but not guaranteed. There is no information available about the identity or number of out-of-database children, however we can infer that these missing children are limited to periods and courts of partial or no database coverage.


Some pertinent features of the network appear in Figure 8. First, no citation arrows are shown, because citations are always from one case to a previous case. Second, multiple observed citations may lead to the same out-of-database document. By design, citations include identifying information, allowing us to match up documents by co-citation even without a complete database. Citations sometimes include a year, allowing us to directly observe the distribution of age of cited laws. Finally, missingness becomes more prevalent with age; in reality this coincides closely with the start of digital record-keeping by most courts in the 1990's.

The citation network of laws has several dynamic elements. It includes a birth process of new documents being written and a death process of them being repealed or expiring by design. The topology between nodes may also change when a document is revised and new citations are added. A document's revision may also affect the probability that the document is cited in the future as well, so revisions can change the underlying model even without changing the topology.

Our method is flexible enough to account for both the dynamic elements of legal documentation, as well as the sampling structure of the CanLII database.

\section{Application of method to Supreme Court database}

We wish to estimate a confidence region for parameters that determine the relative attractiveness of cases to receiving citations.

We do this by creating many simulations of the Supreme Court of Canada (SCC) cases and the citations they receive using the competitive attractiveness model. Each simulation will use randomly generated parameter values from a multivariate uniform prior. Summary statistics will be taken from the results of each simulation. The values of the $N$ parameters of interest and the $M$ statistics from each simulation will be used to create a point in $R^{N+M}$ space.

A Kernel Density Estimator (KDE) is used to create a density function in this space. We take the summary statistics from the actual set of SCC cases investigate the density along the $N$ dimensional hyperplane that matches the observed statistics.

Using the Canadian Law Information Institute's (CanLII) database, we collected 10623 documents produced by the Supreme Court of Canada (SCC). These documents span from the beginning of the SCC in 1867 to August $31^{st}$, 2015. The text processor we designed using the stringr package in R identified 9847 of these documents to be cases.

A case is defined to be a dispute between two parties. One of these parties may be a surrogate for the public, such as ``The Queen'', ``The King'', or simply ``R.'. One or both parties may be a corporation, as identified by the inclusion of ``Company'', ``Co.'', ``Ltd.'', or ``Inc.'' in the name of the party.

In some cases, the opinions of the members of the Supreme Court were not unanimous. In such cases, the judges that were in the minority are listed in a special `dissenting' section of the case document.

Most cases include citations to previous cases as precedents. These citations are identifiable in most or all situations thanks to strict citation naming conventions. However, not every cited case has been recorded digitally or made available by CanLII. Citations to available cases include hyperlinks, making it possible to explore part of the network backwards in time.

Each case also includes a 'cited by' list of the available documents that are known this case. Hyperlinks are included to available reverse citation, but un-digitized work is not listed in these 'reverse citations.

Figure 8 shows a sample, specifically an induced subgraph of the citations between Supreme Court of Canada cases. The vertical axis represents time; recent cases are at the top, and early cases are at the bottom. There are 1500 cases represented in this figure, which is less than 15 percent of all SCC cases. As such, the density of the citations between SCC cases is understated, as only links between cases in the figure are shown. If all 10623 such cases were represented, we would expect approximately 50 times as many lines in the figure.

\begin{figure}[ht]
\begin{center}
\includegraphics[angle=0,scale=0.60]{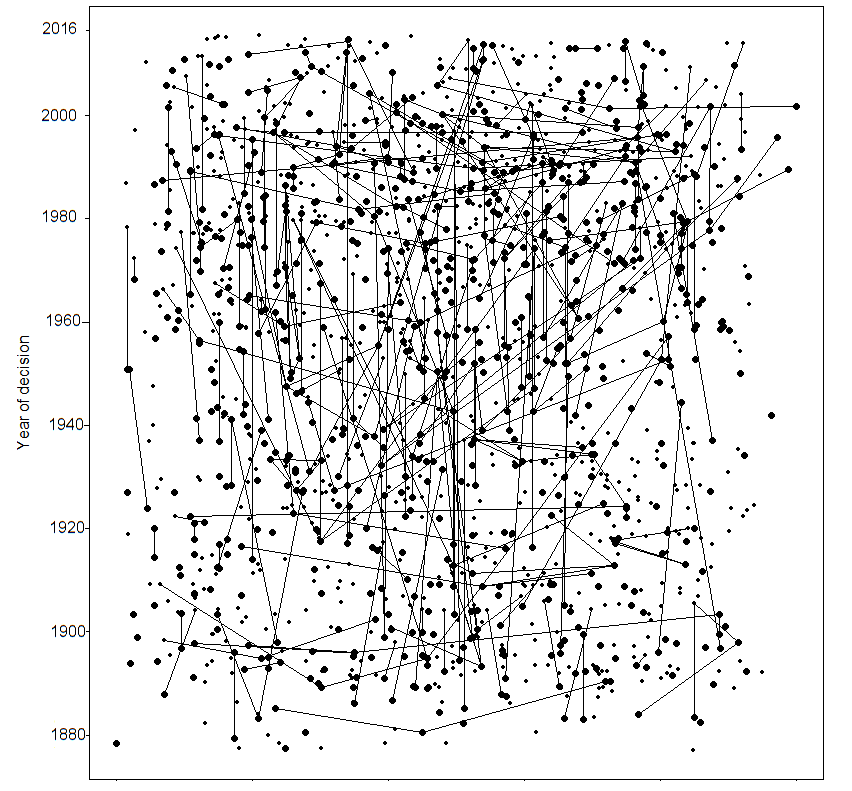}
\end{center}
\vspace{-8.0mm}
\caption{Induced subgraph of the citations between 1500 Supreme Court of Canada cases. Each point represents a case. The size of the point (above a minimum) represents the number of cases from all sources that cite this case. The vertical position of each point represents the year of that case (i.e. the `birth' time). Lines between points represent a citation from the newer case to the older one. The horizontal position of points is arbitrary, and the points have been arranged to minimize the number of crossing lines while preserving vertical position. }
\end{figure}

For the sake of computation speed, time is quantized into steps of 5 years. Table 4 shows the number of Supreme Court cases in each time step, as well as the number of digitized documents citing these cases. The table also includes the proportion of these cases that involve at least one corporation, a dissenting judge, or the crown respectively.

\vspace{5.0mm}
\renewcommand \baselinestretch{1.0}
\begin{table}[htpb]
\begin{center}
\begin{tabular}{l || c | c | c | c | c  }
Period & Cases & Cites Received & Corporate & Crown & Dissent \\
 \hline
1950-4 & 220 & 1185 & 0.32 & 0.25 & 0.52 \\
1955-9 & 287 & 1087 & 0.40 & 0.17 & 0.39 \\
1960-4 & 384 & 1363 & 0.44 & 0.19 & 0.31 \\
1965-9 & 398 & 1675 & 0.43 & 0.21 & 0.30 \\
1970-4 & 425 & 2971 & 0.39 & 0.19 & 0.35 \\
1975-9 & 510 & 4482 & 0.36 & 0.31 & 0.33 \\
1980-4 & 590 & 5672 & 0.29 & 0.38 & 0.20 \\
1985-9 & 462 & 6870 & 0.24 & 0.49 & 0.27 \\
1990-4 & 536 & 9433 & 0.16 & 0.53 & 0.33 \\
1995-9 & 564 & 12373 & 0.24 & 0.52 & 0.36 \\
2000-4 & 460 & 18445 & 0.27 & 0.37 & 0.31 \\
2005-9 & 416 & 21076 & 0.29 & 0.38 & 0.32 \\
2010-4 & 313 & 24947 & 0.26 & 0.44 & 0.26 \\
 \hline
\end{tabular}
\end{center}
\vspace{-5.0mm}
\caption{Summary statistic values for Supreme Court of Canada cases.}
\end{table}

Bell (1975) \nocite{bell1975market} explains the proportion of total sales of a product, or market share, by assigning an attractiveness value $a(s_i)$ to each vendor $i$. In this model, each buyer individually chooses a vendor with probability proportional to this attractiveness. We adapt this attractiveness framework by interpreting each a citation as a purchase, and each case that could be cited as a vendor. According to Table 4, in the 1970-1974 time step, for instance, there were 2971 citations received by SCC cases. The 2971 citations from this time step can refer to any cases, including those that happened long before the year 1970. 

In a given time step, the citation attractiveness of case $i$ is modeled as

\begin{eqnarray}
a(s_i) = \exp\left[ \beta_{irrel}x_{irrel} + \beta_{pa}x_{pa,i} + \beta_{corp}x_{corp,i} + \beta_{crown}x_{crown,i}  + \beta_{dis}x_{dis,i}\right]
\end{eqnarray}

where $x_{irrel,i}$ is the number of time steps since case $i$ was either created or cited, where $x_{corp,i}$, $x_{crown,i}$, and $x_{dis,i}$ are indicator functions of whether case $o$ involved a corporation, the crown, or dissent between judges respectively, and where $x_{pa}$ is the square root of the number of citations that case received in the previous time step.

The parameter $\beta_{irrel}$ represents the tendency of a case to fall into irrelevance over time. The parameter $\beta_{pa}$ represents the tendency towards preferential attachment, whereby a case that receives many citations is likely to be seminal or famous, and therefore is used as a reference more often.

In each of many simulations, data is generated in two ways. Cases and citations before 1950 are generated deterministically from the observed data. For each such case,  $x_{irrel,i}$ is computed from the citation history of each case up to 1950, and $x_{corp,i}$, $x_{crown,i}$, and $x_{dissent,i}$ are taken directly from observation. Cases and citations happening from 1950 onwards are randomly generated. 

Each simulation runs through 13 time steps consisting of the years 1950-4, 1955-9, \ldots , 2010-4. In each time step, cases are created, then citations are assigned, then   $x_{irrel,i}$ and  $x_{irrel,i}$ are updated accordingly. In each simulation, the number of cases created and number of citations assigned in each time step match the true number of Supreme Court cases and citations for that time step. For these cases, $x_{irrel,i} = x_{pa,i} = 0$ upon inception, and $x_{corp,i}$, $x_{crown,i}$, and $x_{dis,i}$ are independently set to 1 with probabilities equal to relevant proportions found in Table 4.   

After case creation, each citation is assigned to an existing case $i$, including those that were created this time step, with probability $a(s_i) / \sum_i{a(s_i)} $ Finally, $x_{irrel,i}$ and $x_{pa,i}$ are updated according to the number of citations they received.

We have selected five statistics to summarize the simulation results. The includes the standard deviation of the distribution of values $\sum_t{C_{t,i}}$, the total number of citations, where $C_{t,i}$ is the number of citations that case $i$ receives in time step $t$.

We use the estimated probability of `going cold'. That is, the chance in any given time step that a citation will receive no new citations, given that it received as least one citation in the time step immediately previous. This is computed by

\begin{eqnarray}
p_{cold,i} = \frac{\sum_{t>1}{1(C_{t-1,i} \geq 1)1(C_{t,i} = 0)}}{1(C_{t-1,i} \geq 1)}
\end{eqnarray}

The other three statistics we use are the estimates of $\gamma_{corp,i}, \gamma_{crown,i}, \gamma_{dis,i}$ in the Poisson family generalized linear model, 

\begin{eqnarray}
\log(\sum_i{C_{t,i}}) = \gamma_{corp,i}x_{corp,i} + \gamma_{crown,i}x_{corp,i} + \gamma_{dis,i}x_{dis,i} + \epsilon, \epsilon \sim Gaussian.
\end{eqnarray}

These summary statistics are also calculated for the actual set of cases and their citation history. Table 5 has the statistic values, before normalization, of the actual cases, as well as the mean and standard deviation for 2000 simulated cases. The prior for the parameters used was the product of independent uniforms $\beta_{irrel} \sim Unif(-0.5,0), \beta_{corp} \sim Unif(-2.0,3.0), \beta_{crown} \sim Unif(-0.5,5.0), \beta_{dis} \sim Unif(0.5,2.0)$

\vspace{5.0mm}
\renewcommand \baselinestretch{1.0}
\begin{table}[htpb]
\begin{center}
\begin{tabular}{l || c | l  }
Statistic & Real Data Value & Mean(SD) of values from simulations\\
 \hline
SD(Total Citations)   & 16.6 & 65.3 (49.2)  \\
``Gone Cold'' probabilitity & 0.743 & 0.440 (0.091) \\
$ \gamma_{corp,i}$ & -0.181 & -0.101 (0.235) \\
$ \gamma_{crown,i}$ & 0.553 & 0.288 (0.336) \\
$ \gamma_{dis,i}$ & 0.373 & -0.131 (0.305) \\
 \hline
\end{tabular}
\end{center}
\vspace{-5.0mm}
\caption{Summary statistic values for Supreme Court of Canada cases, actual and simulated.}
\end{table}

These summary statistics are compared to the statistics of the actual dataset of Supreme Court decisions, a weighted Euclidean distance will be computed between each simulation and the real data, and a weight assigned to each simulation's parameter values based on this distance and on the multivariate normal joint probability distribution. The weighted mean of the simulated parameter values is the Approximate Bayesian Computation estimate of those parameters.


\begin{figure}[ht]
\begin{center}
\includegraphics[angle=0,scale=0.60]{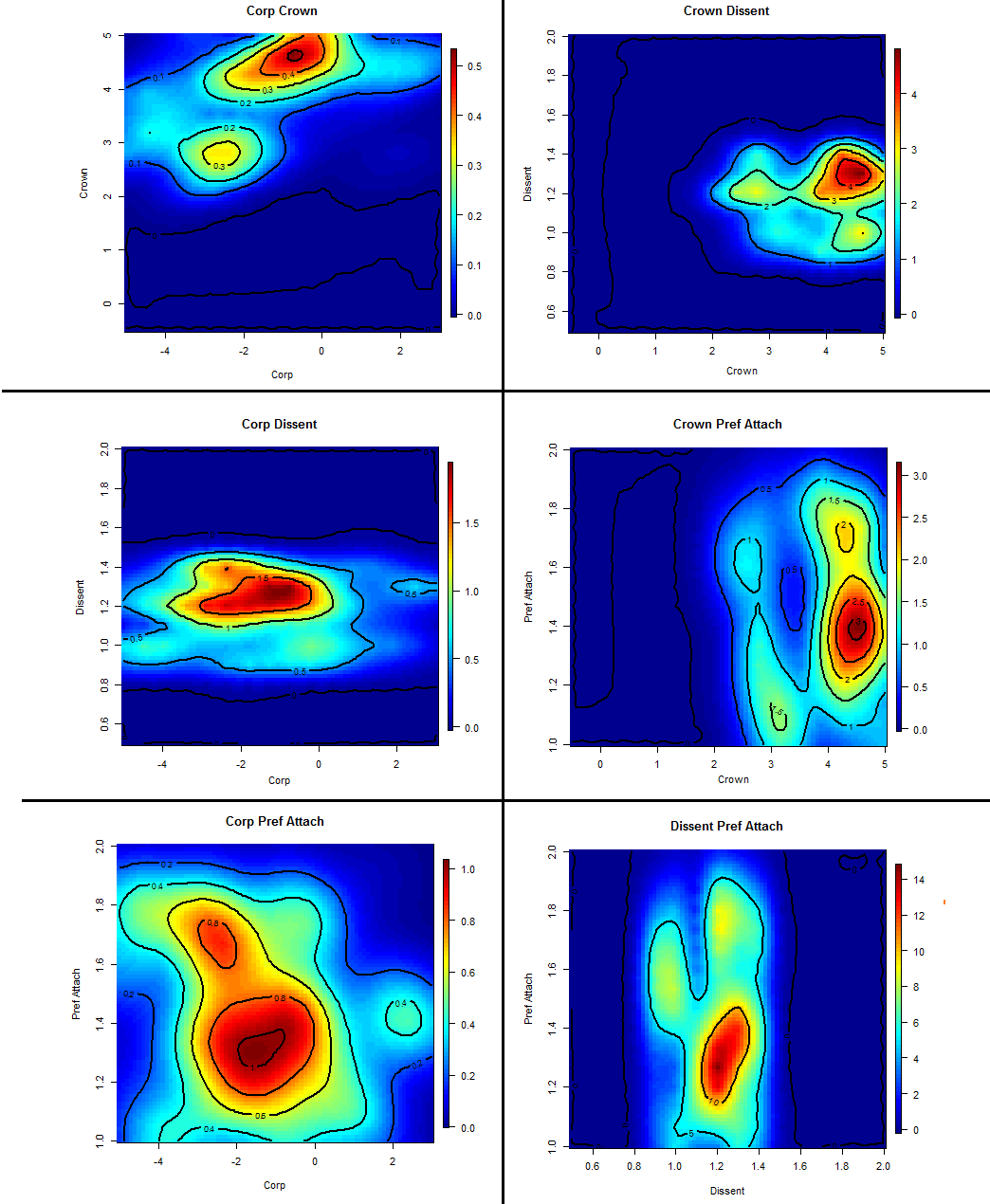}
\end{center}
\vspace{-8.0mm}
\caption{Probability density maps of $\beta_{corp}$,  $\beta_{crown}$, $\beta_{dis}$, and $\beta_{pa}$}
\end{figure}

\section{Discussion}

The proposed method applies the extreme flexibility of approximate Bayesian computation to provide answers to otherwise intractible network inference problems. It is the first attempt, to the author's knowledge, at estimating parameters that govern the generation of a network by using a sample of the network. Futhermore, it is the first known attempt to apply a kernel density estimation method to estimate a parameter distribution for an approximate Bayesian computation. 

The network simulator that was created for this study is adaptable to a wide set of possibilities, including the simulated law citation networks. The output of the simulation program is a detailed data set of each sample node in the population. It provides, in long format for each node, the sampling order, location in social space, the ID of the node, if any, that recruited this one, a number of connections in total, responding connections , and outward connections that were eventually recruited into the sample.

Each line in the output also has application-specific parameters which may be used in the simulation. Consider a case in which the network of interest is a sexual network, such as the one observed in Colorado Springs, USA (Potterat et al. 2002)\nocite{potterat2002risk} or on Likoma Island, Malawi (Helleringer and Kohler 2007)\nocite{helleringer2007sexual}. For cases like this, each node has a sex and a sexual preference. Rather than a single response-to-recruitment probability across the whole population, two rates - one for heterosexual and one for homosexual connections, are generated. Also the connection propensity assigned to each node can be interpreted as a level of sexual promiscuity. The distribution of connection propensity can be made a mixture distribution to account for the subpopulations of sex workers or clients.

As mentioned in Section 4, simulations ideally produce statistics close to the observed sample's statistics. Immediate future work will focus on methods of parameter value generation that optimize the value of a given simulation.

Simulations from a small pilot run across an inclusive parameter space, such as the one used in Section 4, could be combined with a larger main run by weighting simulations by inverse-density. Rather than limiting the parameter space adjustment to a single, manual step, an adaptive system which changes the distributions that parameters are drawn from could be implemented.  Such a system could make cursory estimates of the conditional density and change the shape parameters of the generating distribution to drive future simulations towards the regions of maximal density. This system could also change the scale and location parameters of the generating distribution if a region of non-negligible density was found at the extremes of the parameter space, which may indicate that the global maximum density is not being included in the parameter space yet.

Other avenues of development include using parameter dependency, and computational parallelization. Dependency structures such as a copulas could also be introduced, although the additional value may not be work the added complexity and effort. Parallelization, however, shows immediate potential because the bulk of the computational cost comes from repeating a complex simulation many times over. Parallelization of ABC has gone as far as using graphics processing hardware by Liepe et al. (2010)\nocite{liepe2010abc}, which allows a single desktop computer to run thousands of operations at once. This could be further extended with adaptive rounds by allowing each graphics processing unit (GPU) to communicate interim results to the more general-purpose central processing unit (CPU) acting as a `queen' node.

The proposed method, and the software developed to explore it, have addressed only a handful of protocols of respondent-driven sampling. These protocols have fixed sample sizes, and only edges that lead to recruitment are retained. However, with greater protocol flexibility, practitioners could conduct what-if analyses. That is, before any real sampling is performed, practitioners could estimate from assumed statistics and observed data how the conditional density function and the parameter estimates from it differ under different protocols. For example, one could see on a spatial graph of the conditional density, how much additional uncertainty is introduced by not including edges that leap back into the sample already taken.

Another limitation to be addressed in future work is the sequential nature of the sampling protocol in the code. A `seed' member of the population is only selected by simple random sampling when all sampling-by-recruitment is exhausted. In a practical setting, a wave of multiple subjects is selected by simple random sampling before any recruitment links are followed; additional waves, rather than single subjects, are selected when recruitment links fail to bring in the requisite sample size. The size of these initial cohorts may impact the network graph that arises from the sample, as well as the sampling order. This could be especially problematic if some members of the initial population share a network component and recruitment paths interfere with other.

Uncertainty relating to sampling order is currently addressed in part by timing options in the simulator; users can choose to have recruitment links followed according to a breadth-first search algorithm, or in a fashion where each recruitment link is assigned a random response delay and the links are explored in time-order.

There are some mechanical limitations that need to be addressed before development of the software for this method continues much further. Foremost among these limitations, the network population is stored in a sparse vector-and-matrix system. The edges of a network graph of order $N$ requires $\mathcal{O}(N^2)$ units of memory to store. In relation simulation work in Thompson (2013)\nocite{thompson2013dynamic}, network information is stored as a linked list of node objects. For a maximum number of links per node $M$, storage of the same network graph's edges requires $\mathcal{O}(MN)$ memory, which is effectively $\mathcal{O}(N)$ because $M$ is small compared to $N$ for sparse networks. A linked-list system, although more challenging to program initially, also allows for dynamic network graphs in which nodes undergo birth and death processes.

In future analysis of the law database, we wish to establish criteria for predicting when a legal document is unlikely to be cited ever again. We will do this by modelling the probability that a document will be cited within the next time step (e.g. five years) as a function of the document's intrinsic and network properties.

There are other intrinsic properties year of document, word length of document, whether the document has been revised from its original version, and indication (by key words) that a document involves a criminal offense. Network properties include the presence of a 'child' citation during the previous time step, the presence of a 'child' citation in any of the most recent 2,3, or 4 time steps, the presence of a previous citation in a Supreme Court case.

The model given here is simple, but is ripe for expansion. It treats the documents making the citations as an unconnected pool of identical cases. Citations are obviously made to cases that are relevant to the citing case. A multiplicative factor could be applied to the attractiveness of cases based on the similarity between the citing document and case potentially being cited. 

Similarity could be further established with the key terms attached to most cases. These key terms span from wide categories like ``Criminal law'', ``Constitutional law'', and ``Contract'' to terms used only once like ``Operation of tramway'' and ``Automatic sprinklers''. Employing Approximate Bayesian Computation to a model incorporating key terms could be extremely powerful, but would also involve a realistic simulation of the key terms generated in cases, as well as a text processor that accounted to English and French terms and linguistic changes.

Furthermore, one case may supersede or effectively overwrite another earlier case. Much of the attractiveness of the old case should be transferred to the new case this situation, and this phenomenon could be detected with further network analysis.

Better text mining will be able to better characterize the contents of the document for better prediction. For example, family disputes are not always between family members with the same last name. Cases between parties of very common surnames like Singh and Nguyen can be detected as being family disputes, even when they are not. We were unable to use family disputes as an indicator function for cases because fewer than 2 percent of cases are between parties with the same last name.

Future network analyses may incorporate a wider array of courts and help identify laws with a high probability of irrelevance, and flag them for review and potential repeal. This has civil liberty implications, as well as being a source of political capital for individuals and groups looking to fine easy causes to rally behind and gain popularity. 

Alternatively, this work could identify documents that have high citation potential and flag them as seminal for future law students and researchers. Combining a deeper text analysis with the results of our method may allow for identification of documents with long-term impact from the text alone.

Finally, new cases are frequently examined by the Supreme Court of Canada, and the set of 4 million legal documents in the CanLII database is ever-growing. Even the same analysis could be performed at different times and produce new and interesting results.

Several mechanical improvements to the approximate Bayesian computation - based method of estimating network parameters in Section 6, but little was said about dynamic network possibilities. Future work in the near-and-mid term will include adding the forming and breaking of edges between nodes over time, the drift of nodes in space, and the time-based spreading of infection. For other possibilities, see the network simulation work done in Thompson (2013)\nocite{thompson2013dynamic} which will continue to guide these advancements.

Future work on ABC applications to networks will focus on two related areas: application and distribution.

Before wider dissemination of the ABCN analysis programs, ease of use and speed need to be greatly improved. Currently, simulation is done in a C program and the output, a large csv file, is fed into an R script to analyze results. Ideally, this would all be done as a seamless function, such as one that calls a C-code simulation from R with the help of a dynamic link library (i.e a DLL file).

Computationally, there are two bottlenecks: Simulation, and Kernel Density Estimation. The cost of simulation changes between applications, but every intended application involves a network, and there are common techniques that can be used to improve the speed and memory footprint of networks. These techniques are mentioned in Thompson (2013) and are still under development. The speed of kernel density estimation can be improved by building the analysis around the method outlined in O'Brien et al. (2016) \nocite{o2016fast}, and programmed in the fastKDE package for R. The memory cost of fastKDE is similar to that of traditional kernel density estimation, that is to say, exponentially increasing with the number of distinct summary statistics.

Real network disease datasets like that of Helleringer (2007) are difficult to acquire because of the cost of sampling, and because of concerns for subject privacy. Having a portable software package that is usable by health researchers will allow deeper analyses of these rare and valuable datasets without having to release confidential information beyond its intended range.

The application of our approximate Bayesian computation-based methods in citation analysis have only started. In the law citation example, we only estimated the relative importance of a few features of Supreme Court of Canada cases in determining the ability of these cases to attract citations. Although we considered citations from lesser courts, we did not investigate the citations that these cases from other courts received. We also treated citations as coming from identical cases, when relevance is highly important. In future work, we intend to examine, for example, the attractiveness of corporation-involved cases to future corporation-involved cases. To do so would increase the dimensionality of the kernel density estimation problem, which continues to have the memory cost issue mentioned previously.

%
%
%
%
%

	\bibliographystyle{plain}
	\bibliography{jackthesisref}

\end{document}